\documentclass[lettersize,journal]{IEEEtran}
\usepackage{amsmath,amsfonts}
\usepackage{algorithmic}
\usepackage{algorithm}
\usepackage{array}
\usepackage[caption=false,font=normalsize,labelfont=rm,textfont=rm]{subfig}
\usepackage{textcomp}
\usepackage{stfloats}
\usepackage{url}
\usepackage{verbatim}
\usepackage{graphicx}
\usepackage{cite}
\usepackage{amsmath}
\usepackage[table]{xcolor}
\usepackage{ragged2e} 
\usepackage{booktabs,makecell, multirow, tabularx}
\usepackage{CJKutf8}
\usepackage{booktabs}
\usepackage{multirow}
\usepackage{makecell}
\usepackage{stfloats}
\usepackage{amsmath}
\usepackage[most]{tcolorbox}
\newtcolorbox{observationbox}{
  enhanced,
  colback=green!7,
  colframe=black,
  boxrule=0.8pt,
  arc=3mm,
  boxsep=0mm,
  left=1.8mm,
  right=1.8mm,
  top=1.2mm,
  bottom=1.2mm,
  before skip=4pt,
  after skip=4pt
}

\usepackage{xcolor} 
\usepackage{hyperref} 

\makeatletter
\newcommand{\coloredcite}[1]{{\color{blue}\cite{#1}}}
\makeatother

\hypersetup{
    colorlinks=True,    
    linkcolor=blue,      
    citecolor=blue,    
    filecolor=magenta,  
    urlcolor=magenta,       
    citebordercolor=violet,
  filebordercolor=red,
  linkbordercolor=blue
}

\begin{document}

\title{Life-IQA$^{+}$: Boosting Blind Image Quality Assessment via GCN-enhanced Layer Interaction and Flexible Hierarchical Expert Specialization}

\author{
Long~Tang, Guoquan~Zheng, Huiyu~Duan,~\IEEEmembership{Member,~IEEE},
Jie~Hao, Yucheng~Zhu,\\
Liang~Yuan, Xiongkuo~Min,~\IEEEmembership{Member,~IEEE},
and Guangtao~Zhai,~\IEEEmembership{Fellow,~IEEE}

\thanks{
This work was supported in part by the National Natural Science Foundation of China under Grants 52675031, 62401365, 62225112, 62522116, 62271312, 62571324 and U24A20220; in part by the National Key Research and Development Program of China under Grant 2023YFB4704000; and in part by the China Postdoctoral Science Foundation under Grants BX20250411 and 2025M773473. (Corresponding authors: Liang Yuan, Huiyu Duan)

Long Tang, Jie Hao, Huiyu Duan, Xiongkuo Min are with the School of Integrated Circuits, Shanghai Jiao Tong University, Shanghai 200240, China (e-mail: sjtu8126288@sjtu.edu.cn; sjtu5772066@sjtu.edu.cn;  huiyuduan@sjtu.edu.cn; minxiongkuo@sjtu.edu.cn).

Guoquan Zheng is with the School of Information Science and Technology, Beijing University of Chemical Technology, Beijing 100029, China (e-mail: zhenggq@buct.edu.cn).

Liang Yuan is with the Institute of Cultural and Creative Industry and also with the School of Integrated Circuits, Shanghai Jiao Tong University, Shanghai 200240, China (e-mail: lyuan@sjtu.edu.cn).

Yucheng Zhu is with the Institute of Cultural and Creative Industry, Shanghai Jiao Tong University, Shanghai 200240, China (e-mail: zyc420@sjtu.edu.cn).


Guangtao Zhai is with the School of Computer Science and Technology, Xinjiang University, Urumqi 830049, China, also with Shanghai Artificial Intelligence Laboratory, Shanghai 200003, China, and also with the Institute of Image Communication and Network Engineering, Shanghai Jiao Tong University, Shanghai 200240, China (e-mail: zhaiguangtao@sjtu.edu.cn).
}
}



\maketitle

\begin{abstract}

Blind image quality assessment (BIQA) is essential for evaluating and optimizing visual experience.
Existing BIQA methods typically aggregate features from multiple backbone stages without accounting for their unequal contributions to quality prediction. Furthermore, uniformly supervising all experts may constrain their flexibility and limit the diversity of the learned quality representations.
To address these issues, we present \textbf{Life-IQA$^{+}$}, a quality representation decoding framework that combines graph convolutional network (GCN) enhanced \underline{l}ayer \underline{i}nteraction and \underline{f}lexible hierarchical \underline{e}xpert specialization.
Specifically, the layer interaction module constructs quality queries from GCN-refined deepest-stage features and uses penultimate-stage features as keys and values, thereby enabling cross-layer representation interaction via the cross-attention mechanism. Building on these representations, we further introduce a flexible hierarchical mixture-of-experts (MoE) architecture comprising a densely activated shared expert layer and a sparsely activated routed expert layer. To enhance the representation diversity and ability of experts, we apply auxiliary classification supervision to a subset of shared experts using distortion-type labels for synthetic datasets and discretized quality-level labels for authentic datasets.
Experiments on synthetic and authentic distortion benchmarks, together with controlled ablations of expert configurations and supervision strategies, show that Life-IQA$^{+}$ achieves competitive prediction accuracy with a favorable balance between accuracy and computational cost.
The code is available at: \href{https://github.com/TANGLONG2/Life-IQA-PlusPlus}{\texttt{Life-IQA$^{+}$}}.

\end{abstract}

\begin{IEEEkeywords}
Blind image quality assessment, graph convolutional networks, layer interaction, hierarchical mixture-of-experts, expert specialization, multi-task learning.
\end{IEEEkeywords}

\section{Introduction}
\label{sec:intro}

\IEEEPARstart{I}{mage} quality assessment (IQA) aims to predict the perceptual quality of images aligned with human judgments \coloredcite{10109108,11092710,11093358,11175315,10795128,10056825,9961939}. Among different IQA paradigms, blind image quality assessment (BIQA) estimates image quality without access to pristine reference images \coloredcite{10960664,10647859,10833846,10463077,10462930}, making it particularly suitable for practical scenarios where reference images are typically unavailable. Existing BIQA methods extract multi-scale features from distorted images using pretrained backbone networks and aggregate them for quality prediction \coloredcite{9156687,9817377,9121773,10323206,10036133}. This strategy assumes that features from different backbone stages, with varying spatial resolutions and semantic abstraction levels, can provide complementary information for quality prediction. 
Specifically, given an input image with height $H$ and width $W$, a typical backbone produces multi-stage features denoted as $stage1\in \mathbb{R}^{\frac{H}{4}\times \frac{W}{4}\times C_1}$, $stage2\in \mathbb{R}^{\frac{H}{8}\times \frac{W}{8}\times C_2}$, $stage3\in \mathbb{R}^{\frac{H}{16}\times \frac{W}{16}\times C_3}$, and $stage4\in \mathbb{R}^{\frac{H}{32}\times \frac{W}{32}\times C_4}$, where $C_i$ is the channel dimension of the $i$-th stage. Shallow features ($stage1$, $stage2$) primarily capture local details and texture information, whereas deep features ($stage3$, $stage4$) encode global and semantic information. Despite their complementary roles in perceiving features, the relative contributions to quality prediction remain underexplored.


\begin{figure}[t]
\label{sec:intro}
\centering
\includegraphics[height=5cm]{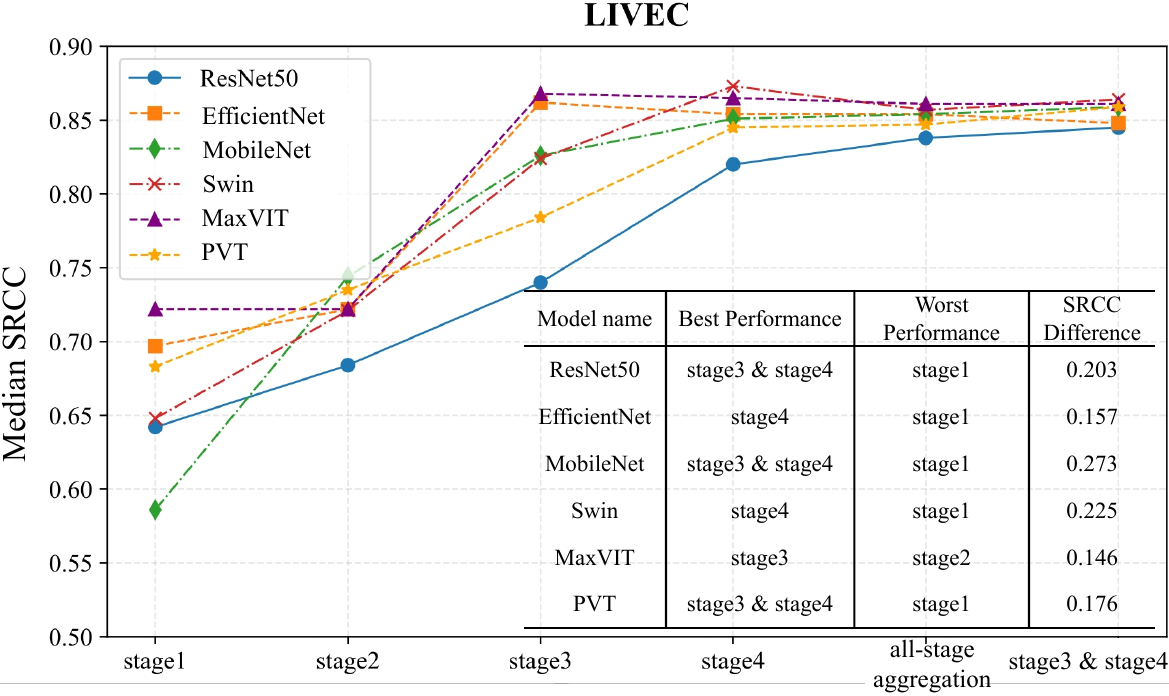}
\caption{Performance of each stage and fused features from different pretrained models on LIVEC. $Stage3$, $stage4$, and their fusion perform better.}
\vspace{-4mm}
\label{fig_1}
\end{figure}

To quantify the actual contribution of features from different stages to the BIQA, six different pretrained backbones are evaluated on the authentic distortion dataset LIVEC \coloredcite{7327186}. These backbones include three convolutional neural
networks (CNNs), \textit{i.e.}, ResNet50 \coloredcite{he2015deepresiduallearningimage}, EfficientNet \coloredcite{tan2020efficientnetrethinkingmodelscaling}, MobileNet \coloredcite{howard2019searchingmobilenetv3} and three Transformer models, \textit{i.e.}, Swin Transformer \coloredcite{liu2021swintransformerhierarchicalvision}, MaxViT \coloredcite{tu2022maxvitmultiaxisvisiontransformer}, PVT \coloredcite{wang2021pyramidvisiontransformerversatile}. To separately investigate the quality-representation capacity of the features from different layers, the image features from each stage are directly appended with a simple prediction head, which consists of a global average pooling layer (GAP) and a linear regression layer. We also adopt two simple feature pyramid fusion methods for comparison, including an all-stage aggregation method, and a $stage3$ \& $stage4$ fusion scheme. These two methods first align the spatial resolution and channel dimension of the features from different stages, then adopt the same simple prediction head to aggregate. As shown in Fig.~\ref{fig_1}, the shallow features ($stage1$, $stage2$) consistently achieve lower median SRCC values than the deep features ($stage3$, $stage4$) across all six backbones. Moreover, $stage3$ \& $stage4$ achieves performance comparable to all stage aggregation. 
This suggests that directly fusing all stage features is not always beneficial, as shallow features dominated by low-level details may introduce redundancy or noise into quality regression, especially with limited BIQA training data.

Based on the aforementioned analysis, rather than directly adopting the conventional strategy of aggregating all multi-scale features, we focus on achieving efficient interaction among deeper, semantically richer features. To this end, we first propose \textbf{Life-IQA}, an effective quality feature decoding framework built upon \textbf{GCN-enhanced Layer Interaction} and \textbf{MoE-based Feature Decoupling}. In Life-IQA, the deepest-stage features, enhanced through a graph convolutional network (GCN), serve as query tokens that carry high-level quality semantics, while the penultimate-stage features act as keys and values, providing structured local evidence, then the cross-attention-based interaction is performed. Furthermore, a mixture-of-expert (MoE) module decouples the fused representations, allowing different experts to specialize in particular distortion types or quality dimensions. This design yields a dynamic, data-efficient decoding mechanism tailored to BIQA characteristics. However, the original MoE uses only sparsely routed experts and lacks a dedicated pathway for modeling quality information shared across samples.


\begin{figure*}[t]
\label{sec:intro}
\centering
\includegraphics[width=7in]{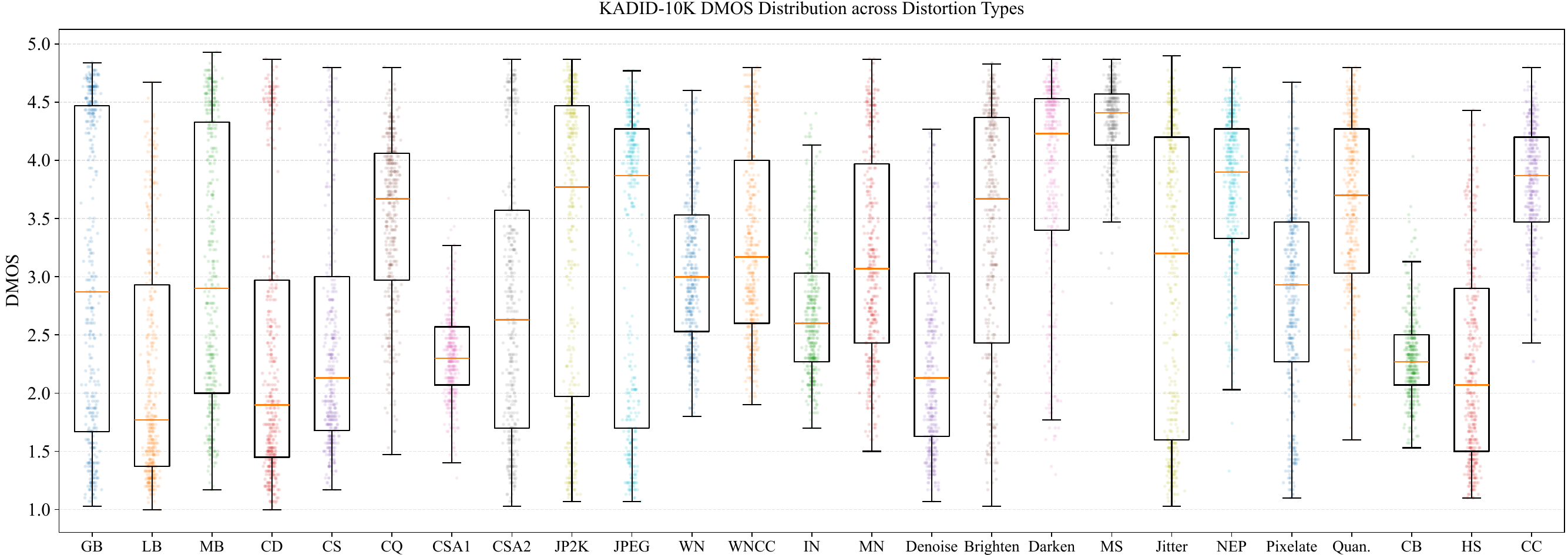}
\caption{Distribution of Difference Mean Opinion Scores (DMOS) across the 25 distortion types in KADID-10K. Different distortion types exhibit substantially overlapping DMOS distributions, while samples within the same distortion type span multiple quality levels, indicating a non-one-to-one correspondence between distortion classification labels and quality regression labels. }

\label{fig_2}
\end{figure*}

A natural extension is to introduce densely activated shared experts in parallel with the original routed experts, following the shared expert organization adopted in DeepSeekMoE \coloredcite{dai2024deepseekmoeultimateexpertspecialization}. In this parallel MoE, the shared experts capture perceptual cues that recur across samples, while the routed experts preserve input-dependent specialization through sparse activation. Although this design specifies the roles of shared and routed experts, how shared experts should be guided to learn quality-relevant common knowledge remains underexplored in BIQA. 

Some works have proposed multi-task learning-based BIQA methods\coloredcite{7351311,9337868,10204315,wu2023qalign,Duan_2025_CVPR}, which additionally consider learning distortion-type labels for synthetically distorted datasets or quality-level labels for authentically distorted datasets. However, these methods have not jointly consider both distortion-type and quality-level supervision during learning quality score prediction, and have not consider using different experts to learn different attributes. As shown in Fig.~\ref{fig_2}, different distortion types exhibit substantially overlapping difference mean opinion score (DMOS) distributions on KADID-10K, while images within the same distortion type span a broad range of quality scores. Moreover, different distortion types exhibit significantly different influences on quality levels, further leading to different quality score distributions. This suggests that effectively leveraging MoE for BIQA requires an expert learning strategy that incorporates distortion-related characteristics or coarse quality levels while preserving flexibility for quality regression.

To this end, we propose \textbf{Life-IQA$^{+}$}, a BIQA framework that integrates
\textbf{GCN-enhanced
\underline{L}ayer
\underline{I}nteraction}
and \textbf{\underline{F}lexible Hierarchical \underline{E}xpert Specialization}. Specifically, Life-IQA$^{+}$ contains both densely-shared and sparsely-routed experts, which are organized sequentially rather than in a single parallel stage. The shared expert layer learns reusable quality representations, with auxiliary classification supervision applied only to a subset of experts to learn distortion-type and quality-level attributes, respectively. The routed expert layer then dynamically activates a small subset of experts for each quality token to perform adaptive quality-representation refinement. This selective and sequential design incorporates auxiliary information into dedicated expert paths rather than imposing uniform constraints on all experts. Overall, the main contributions are summarized as follows.

\begin{itemize}

\item We develop a selective deep-layer decoding strategy for BIQA. The deepest-stage features generate content-guided quality queries, while the penultimate-stage features provide complementary keys and values for cross-attention.

\item We introduce a lightweight GCN-based query modeling mechanism that explicitly captures the topology and dependencies among quality queries before cross-attention, producing structured queries for subsequent feature aggregation.

\item We design a flexible hierarchical MoE that places a densely activated shared expert layer before a sparse routed expert layer, enabling reusable quality modeling followed by input-dependent residual refinement.

\item We propose a selective auxiliary supervision strategy that applies distortion-type classification for synthetic datasets and quality-level classification for authentic datasets to only a subset of shared experts.

\item Extensive experiments on seven BIQA benchmarks demonstrate competitive prediction accuracy and favorable computational efficiency. Compared with Life-IQA, Life-IQA$^{+}$ reduces the parameter count from 95M to 45M while improving the average performance.

\end{itemize}
A preliminary version of this work, Life-IQA \coloredcite{Tang_2026_CVPR}, was published as a CVPR highlight. In this journal extension, we make several substantial improvements. (1) We extend the original MoE-based feature decoupling module into a flexible hierarchical MoE framework, where shared experts provide degradation-aware residual enhancement and sparse routed experts perform sample-adaptive quality refinement. (2) We introduce auxiliary-supervised expert learning for both synthetic and authentic BIQA datasets, using distortion type classification and quality level classification, respectively. Unlike conventional multi-task learning methods, the auxiliary supervision is embedded into a subset of shared experts rather than used as an external prediction branch. (3) We provide more comprehensive analysis and experiments, including comparisons with conventional multi-task learning, different MoE designs, expert activation visualization, routing behavior analysis, cross-dataset evaluation, data-efficient evaluation, efficiency comparison, and detailed ablations. Compared with the conference version, this journal extension provides a more complete expert-learning framework for BIQA.

\section{Related Work}
\subsection{Learning based Image Quality Assessment}
Learning-based BIQA methods learn quality-sensitive representations directly from distorted images and map them to subjective quality scores. Their performance depends not only on the representation capacity of the backbone, but also on how features at different spatial scales and semantic levels are aggregated. Existing deep BIQA methods have mainly developed along two architectural lines: CNN-based models and Transformer-based models.

CNN-based BIQA methods commonly exploit hierarchical feature maps produced by convolutional backbones. Su \textit{et al.} \coloredcite{9156687} used ResNet50 to extract multi-scale features and employed semantic information to adapt the quality regressor. Lan \textit{et al.} \coloredcite{10092271} combined a generative adversarial network with EfficientNet to capture local distortion details and high-level semantics, respectively, and fused them through a feature pyramid. Although these methods demonstrate the value of hierarchical representations, their fusion mainly relies on concatenation or weighted aggregation, without explicitly selecting information according to its relevance to quality prediction.

Following the development of the Vision Transformer (ViT) \coloredcite{dosovitskiy2021imageworth16x16words}, Transformer architectures have also been introduced into BIQA \coloredcite{10478301,9506075}. Golestaneh \textit{et al.} \coloredcite{9706735} combined CNN-based local feature extraction with Transformer modeling and introduced relative ranking and self-consistency objectives. Ke \textit{et al.} \coloredcite{9710973} employed hash-based spatial embedding to process images at their native resolutions and aspect ratios. These methods mainly use Transformers as encoders for feature extraction and fusion, while query-based decoder aggregation has received less attention. DEIQT \coloredcite{qin2023dataefficientimagequalityassessment} is one of the few BIQA methods to adopt an encoder-decoder architecture. However, its queries are generated from a single CLS token, which restricts their information source and makes the decoding process dependent on the representation produced by the upstream encoder. This limitation motivates decoder designs that construct queries from richer deep features and selectively aggregate complementary information across backbone stages.

\begin{figure*}[!t]
\label{sec:intro}
\centering
\includegraphics[width=7in]{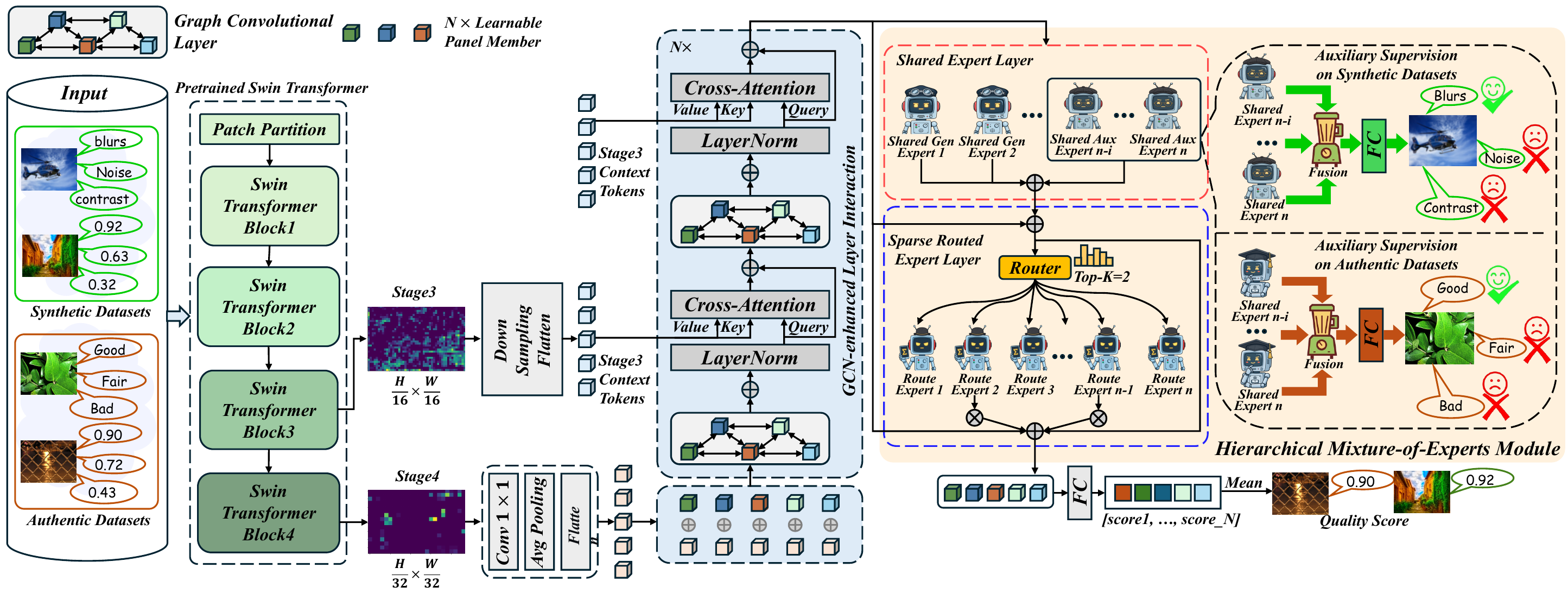}
\caption{Overall framework of the proposed Life-IQA$^{+}$. 
A pretrained Swin Transformer is adopted to extract $stage3$ and $stage4$ features from the input image. 
The GCN-enhanced layer interaction module generates quality-oriented queries from the $stage4$ features and aggregates complementary contextual information from the $stage3$ features via cross-attention. 
The hierarchical MoE module contains a shared expert layer and a sparse routed expert layer, where auxiliary supervision guides shared experts to learn degradation-aware knowledge and the routed experts dynamically collaborate for final quality perception. 
The token-wise quality predictions are averaged to produce the final image quality score.}

\label{fig_3}
\end{figure*}

\subsection{Mixture of Experts}
MoE is a conditional computation architecture that increases model capacity through input-dependent expert weighting. A standard MoE layer contains a set of expert networks and a trainable gating network that determines their contributions for each input. Jacobs \textit{et al.} \coloredcite{6797059} introduced the adaptive mixture of local experts, allowing individual experts to specialize in different regions of the input space. Eigen \textit{et al.} \coloredcite{eigen2014learningfactoredrepresentationsdeep} later extended the conventional single-layer MoE into Deep MoE by stacking multiple expert and gating layers. Each layer operates on the mixture representation produced by the preceding layer, allowing an input to be modeled through different combinations of experts across depth and supporting factorized specialization at different stages.

With the development of large-scale Transformers, sparse MoE has been widely used to replace dense feed-forward networks with sparsely activated expert layers \coloredcite{lepikhin2020gshardscalinggiantmodels}. For example, Mixtral $8\times7$B \coloredcite{jiang2024mixtralexperts} employs eight feed-forward experts in each MoE layer and activates two experts for every token. Recent studies have also reconsidered how expert capacity should be organized. DeepSeekMoE \coloredcite{dai2024deepseekmoeultimateexpertspecialization} introduces fine-grained expert segmentation and shared expert isolation. Its shared experts are activated for every token to capture common knowledge, while only a sparse subset of routed experts is selected for specialized processing; both outputs are combined within the same MoE layer. MoE has also been studied in vision models through vision-specific expert designs and patch-level routing \coloredcite{11226953,zhou2023brainformers,10.5555/3618408.3618649}. However, most existing MoE architectures are developed for large-scale language modeling or generic visual recognition. Their expert organization and supervision strategies do not directly address BIQA, which relies on continuous quality regression and limited subjective annotations. Life-IQA$^{+}$ instead organizes shared and routed experts into successive processing stages and applies auxiliary supervision only to selected shared experts.

\subsection{Multi-Task Learning}
Multi-Task Learning (MTL) aims to let a single model learn multiple related tasks jointly, aiming to improve the model performance by using the correlation between different tasks. This joint learning approach provides more image degradation information for quality score regression BIQA task. Kang \textit{et al.} \coloredcite{7351311} proposed a compact CNN to simultaneously estimate image quality and identify distortion types by sharing convolutional features. Sun \textit{et al.} \coloredcite{9720138} used a graph neural network to distinguish image distortion type and severity. 

These methods commonly adopt a shared-backbone design, where quality regression and auxiliary prediction are performed on the same extracted representation, followed by independent prediction heads for different supervision signals. In this way, multiple tasks rely entirely on the same set of backbone features, which may introduce task interference or even negative transfer. Specifically, although distortion type recognition provides degradation-aware information for quality assessment, its objective is not fully consistent with quality regression. Large perceptual quality variations may exist within the same distortion category, and different distortion categories may also lead to similar subjective quality degradation. In contrast, soft parameter sharing allows different tasks to have relatively independent parameters or branches, and realizes information sharing through feature interaction or regular constraints, which can theoretically alleviate the task conflict caused by hard sharing. For example, Misra \textit{et al.} \coloredcite{7780802} proposed a cross-stitch unit network to find the best shared representation in multi-task learning for the combination of semantic segmentation and surface normal prediction, target detection and attribute prediction tasks. However, the primary objective of BIQA remains quality score regression, and BIQA datasets are usually limited in scale. Therefore, soft parameter sharing methods are not commonly adopted in BIQA. In this paper, we design a hierarchical MoE network to achieve more flexible feature learning.





\section{The Proposed Method}
\subsection{Overall Architecture}

The overall framework of our method is illustrated in Fig.~\ref{fig_3}, which consists of two key components. The first is a GCN-enhanced layer interaction, responsible for selective cross level feature fusion. This layer interaction generates quality queries from the deepest features and aggregates key information from the penultimate deep features through cross-attention mechanisms. The second component is a hierarchical MoE module, which guides experts to perform specialized learning for BIQA. 
The first shared expert layer densely activates all experts, with a subset additionally guided by auxiliary IQA supervision to learn common quality cues. The resulting representation is then processed by a sparse routed expert layer for dynamic expert selection, while both layers refine the original features through residual connections.


\subsection{GCN-enhanced Layer Interaction}
The GCN-enhanced layer interaction module selectively combines the two deepest backbone stages. The $stage4$ representation provides compact global context, whereas $stage3$ retains finer spatial information. We use $stage4$ to construct image-conditioned quality queries and $stage3$ as the source of keys and values for cross-attention. The module first initializes the queries with $stage4$, then models dependencies among them with a GCN, and finally retrieves complementary information from $stage3$.

A standard Transformer decoder employs a set of learnable embeddings as initial queries $Q_\text{init}\in \mathbb{R}^{N\times D}$, where $N$ is the number of query tokens and $D$ is the channel dimension, set to 384. Since $Q_\text{init}$ is learned independently of the input image, it lacks input-dependent visual context at initialization. We therefore condition these queries on the global representation from $stage4$. Specifically, a $1\times1$ convolutional layer ($conv1$) first reduces the channel dimension of $stage4$ to $D$, and global average pooling (GAP) produces a context vector $v_\text{global}\in \mathbb{R}^{D}$. The vector is then expanded to the query length $N$ and added element-wise to $Q_\text{init}$, yielding the guided initial queries $Q_\text{init}^{'}\in \mathbb{R}^{N\times {D}}$:
\begin{equation}
Q_\text{init}^{'}\ = \text{expand}\left( \text{GAP}\left( conv1\left( stage4 \right) \right) \right) \ +\ Q_\text{init}.
  \label{eq:important}
\end{equation}
This addition preserves the identity of each learnable query while supplying the same input-dependent global context to the entire query set.

Before retrieving information from $stage3$, the initialized queries need to exchange information with one another. A standard Transformer decoder models this interaction through self-attention, which can be written as:
\begin{equation}
\text{Attention}\left( Z_{i+1} \right)  = \text{softmax} \left( \frac{Q_{i+1}K_{i+1}^{T}}{\sqrt{d_k}} \right) V_{i+1}, 
\label{eq:important}
\end{equation}
where $d_k$ is the key dimension. The query, key, and value matrices $Q_{i+1}$, $K_{i+1}$, and $V_{i+1}$ are obtained from the previous-layer representation $Z_i$ through independent linear projections:
\begin{equation}
Q_{i+1}=f_1\left( Z_i \right) ,\ K_{i+1}=f_2\left( Z_i \right) ,\ V_{i+1}=f_3\left( Z_i \right).
\label{eq:important}
\end{equation}
Here, $f_i$ denotes a linear transformation applied independently to each token. Although the subsequent dot-product attention allows the tokens to interact, their relations are induced implicitly from the projected features. We instead use a learnable graph topology to establish explicit propagation paths among the quality queries.

Specifically, we replace query self-attention with a three-layer GCN. The guided query sequence $Q_{\text{init}}^{'}\in \mathbb{R}^{N\times {D}}$ is treated as a graph containing $N$ nodes, each represented by a $D$-dimensional feature. A learnable adjacency matrix $A_i\in \mathbb{R}^{N\times N}$ defines the graph structure of the $i$-th GCN layer and models dependencies among the query nodes. The layer-wise propagation is formulated as:
\begin{align}
\mathcal{Q}_1 &= f_{\text{input}}(Q_{\text{init}}^{'}, A_{1}) = \sigma(A_{1} \cdot Q_{\text{init}}^{'} \cdot W_0), \\
\mathcal{Q}_2 &= f_{\text{hidden}}(H_1, A_{2}) = \sigma(A_{2} \cdot H_1 \cdot W_1), \\
\mathcal{Q}_3 &= f_{\text{output}}(H_2, A_{3}) = A_{3} \cdot H_2 \cdot W_2,
\end{align}
where $H_1$ and $H_2$ are the node representations produced by the first and second GCN layers, corresponding to $\mathcal{Q}_1$ and $\mathcal{Q}_2$, respectively. The matrices $W_0$, $W_1$, and $W_2$ are trainable feature transformations, and $\sigma$ is a non-linear activation function. 

After three propagation layers, $\mathcal{Q}_3$ contains information aggregated according to the learned query dependencies and is used as the query sequence for cross-attention.
The GCN-refined queries are further complemented by the $stage3$ representation. Cross-attention allows the query sequence and the key/value sequence to have different lengths, so features with different spatial resolutions can interact without direct alignment. However, flattening the full $stage3$ feature map would result in a long sequence and high attention cost. To reduce this cost, partition average pooling \coloredcite{10222634} is applied to the $stage3$ feature map. Specifically, $stage3$ is divided into an $N\times N$ grid, and average pooling is applied within each region. The pooled regional features form a compact sequence $S\in \mathbb{R}^{N^2\times D}$, in which each token represents one spatial region. The cross-attention module uses $\mathcal{Q}_3$ as the query and generates the keys and values from $S$. Each quality query can thus retrieve relevant information from $stage3$, while the GCN has already modeled dependencies across the query set. The resulting quality tokens combine the global context supplied by $stage4$ with the regional information from $stage3$ and are passed to the subsequent expert module.


\begin{figure}[t]
\centering
\includegraphics[width=3.5in]{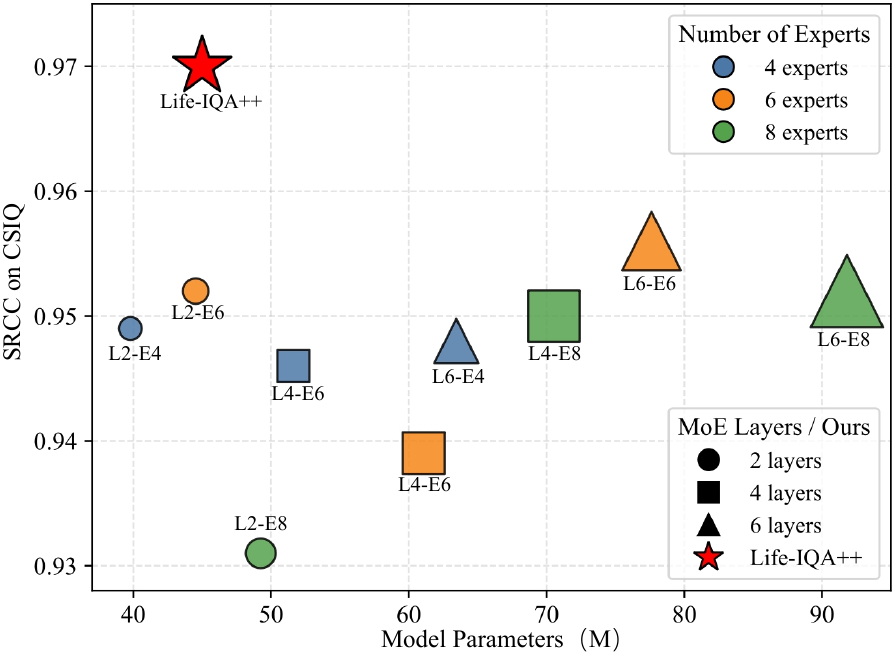}
\caption{Effects of MoE depth and expert count on CSIQ. FFNs in 2, 4, or 6 decoder layers are replaced by sparse MoE layers with 4, 6, or 8 experts. Marker shape and color denote MoE depth and expert count, respectively. Larger configurations increase model size without consistent SRCC gains.}
\vspace{-5mm}
\label{fig_4}
\end{figure}
\subsection{MoE-based Feature Decoupling}
The application of MoE in BIQA remains underexplored. A prevalent design choice for MoE follows the Transformer paradigm, where standard FFNs are replaced by sparse expert layers across multiple blocks. However, BIQA datasets are usually limited by subjective annotations. Therefore, stacking MoE layers exacerbates parameter redundancy and overfitting risks for BIQA. To examine this issue, we conduct a preliminary experiment on CSIQ \coloredcite{larson2010most}. Following the feature preparation process in the GCN-enhanced layer interaction module, we extract the $stage3$ and $stage4$ features from the pretrained Swin Transformer. The $stage4$ feature is projected by a $1 \times 1$ convolution and average pooling to form query tokens, while the $stage3$ feature is used as keys and values through the same partition pooling operation in the Transformer decoder. Then, we replace the FFNs in different decoder blocks with sparse MoE layers and vary both the number of MoE layers and the number of experts. To improve the reliability of the comparison, each configuration is independently run 10 times, and the median performance is reported.
As shown in Fig.~\ref{fig_4}, we replace the FFNs in 2, 4, or 6 Transformer decoder layers with sparse MoE layers and set the number of experts in each MoE layer to 4, 6, or 8.
These configurations cover a wide range of model sizes, from about 40M parameters in the compact setting to more than 90M parameters in the largest setting. 
The analysis results are presented in Fig.~\ref{fig_4}, yielding the following observation.

\begin{figure*}[t]
  \centering
  \includegraphics[width=7in]{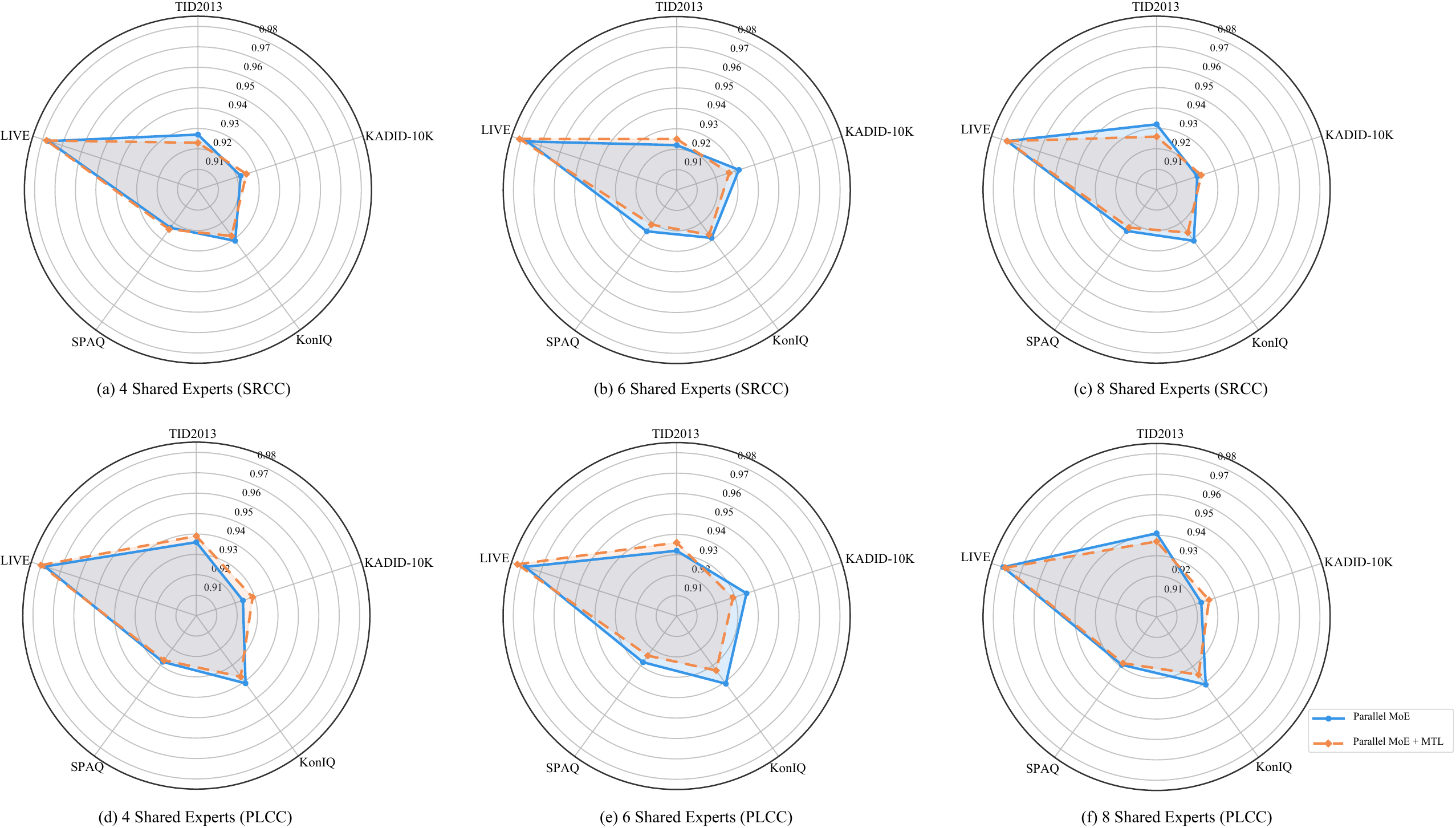}
\caption{Effects of auxiliary supervision on Parallel MoE across five BIQA datasets. Parallel MoE is compared with a multi-task learning (MTL) variant that applies auxiliary classification supervision to the outputs of all shared experts. The auxiliary task uses distortion-type labels for synthetic distortion datasets and quality-level labels obtained by discretizing quality scores into five classes for authentic distortion datasets. Models with 4, 6, and 8 shared experts are evaluated with all other components fixed, and the median SRCC and PLCC values over ten independent runs are reported. Auxiliary supervision does not consistently improve performance across datasets or numbers of shared experts.}
  \label{fig_5}
\vspace*{-5pt}
\end{figure*}

\begin{observationbox}
\textbf{Observation-1.}  Increasing the number of MoE layers or experts consistently enlarges the model, but prediction performance does not improve with the added capacity. Across the tested configurations, SRCC fluctuates as the model becomes deeper or wider, and several compact configurations match or outperform their larger counterparts. These results show that increasing the number of sparse expert paths does not guarantee better BIQA performance. 
\end{observationbox}

\noindent\textbf{MoE Design in Life-IQA.} Motivated by Observation-1, Life-IQA avoids replacing the FFNs across multiple decoder blocks. Instead, Life-IQA places a single MoE module after the GCN-enhanced layer interaction module. This placement allows the experts to refine quality representations that already integrate information from different backbone stages. Within the MoE module, a router dynamically selects a sparse subset of experts for each quality token, allowing different experts to specialize in input-dependent quality patterns without repeatedly expanding MoE layers throughout the decoder. The aggregated expert output is then combined with a scaled residual of the original quality token, preserving the input representation during sparse expert transformation. \mbox{Let $X=\{x_i\}_{i=1}^{N}\in\mathbb{R}^{N\times D}$} denote the refined quality tokens and $\mathcal{R}=\{R_m\}_{m=1}^{M_r}$ denote the routed experts. For each token $x_i$, Life-IQA computes the sparse routing weights and aggregates the selected expert outputs as:
\begin{align}
\mathbf{g}_i &= \operatorname{Softmax}\!\left(\operatorname{TopK}\!\left(G_r(x_i),K\right)\right)
\label{eq:life_iqa_routing},\\
\Delta_i^r &= \sum_{m=1}^{M_r}g_i^m R_m(x_i),
\qquad
x_i^r=\Delta_i^r+\gamma x_i,
\label{eq:life_iqa_sparse_moe}
\end{align}
where $G_r(\cdot)$ is the router, and $\operatorname{TopK}(\cdot,K)$ retains the $K$ largest routing logits while masking the others before softmax normalization. The residual coefficient $\gamma$ preserves the original quality representation.

\subsection{Hierarchical Mixture-of-Experts and Multi-Task Learning }

\noindent\textbf{A Preliminary Analysis of shared expert Guidance.}
Although the sparse MoE in Life-IQA separates input-dependent quality patterns, different routed experts may still learn similar quality cues, such as structural degradation and detail loss. A natural extension is to introduce densely activated shared experts in parallel with the routed experts, following the shared expert organization of DeepSeekMoE \coloredcite{dai2024deepseekmoeultimateexpertspecialization}. Existing multi-task BIQA methods use auxiliary classification tasks to supplement scalar quality regression with degradation or quality level information, thereby providing richer supervision for representation learning. Whether such auxiliary supervision can be directly transferred to the shared experts of a parallel MoE remains unclear. 

Following this practice, we construct a multi-task learning (MTL) variant of the parallel MoE, in which all shared experts are jointly optimized with quality regression and the corresponding auxiliary classification objective. To examine the effect of auxiliary supervision on shared expert learning, we compare the parallel MoE with its MTL variant across BIQA datasets and shared expert capacities. The analysis results are presented in Fig.~\ref{fig_5}, revealing the following observation.

\begin{observationbox}\textbf{Observation-2.} Across the tested synthetic and authentic datasets, applying the corresponding auxiliary labels to all shared experts produces inconsistent effects on the parallel MoE. It improves quality prediction in some configurations, but brings little benefit or even degrades performance in others. Therefore, we argue that directly supervising the entire shared expert branch with these auxiliary labels cannot consistently improve quality-aware representation learning.\end{observationbox}

\noindent\textbf{Optimization Analysis of Auxiliary Guidance.}
Let $\theta_s$ denote the parameters of the shared experts, and let $\mathcal{L}_{q}$ and $\mathcal{L}_{c}$ denote the quality regression loss and auxiliary classification loss, respectively. The auxiliary label $c$ is defined as:
\begin{equation}
c=
\begin{cases}
d, & \text{for synthetic distortion datasets},\\
\mathcal{Q}(y), & \text{for authentic distortion datasets},
\end{cases}
\end{equation}
where $d$ is the distortion type, $y$ is the continuous quality score, and $\mathcal{Q}(\cdot)$ denotes score discretization.

In either case, $c$ provides only a partial description of perceptual quality. For synthetic distortions, images with the same distortion type may have different quality scores, while different distortion types may produce overlapping score distributions. For authentic distortions, discretization does not preserve score differences within the same quality level. This ambiguity can be expressed as:
\begin{equation}
\operatorname{Var}(y\mid c)>0.
\end{equation}
which indicates that the auxiliary label cannot uniquely determine the continuous quality score. When all shared experts are jointly optimized by both objectives, the joint loss is:
\begin{equation}
\mathcal{L}=\mathcal{L}_{q}+\lambda_{c}\mathcal{L}_{c},
\end{equation}
where $\lambda_{c}$ controls the contribution of the auxiliary classification loss. The gradients of the two objectives with respect to the shared expert parameters are:
\begin{equation}
\mathbf{g}_{q}=\nabla_{\theta_s}\mathcal{L}_{q},
\qquad
\mathbf{g}_{c}
=
\nabla_{\theta_s}\mathcal{L}_{c}.
\end{equation}
Accordingly, under one gradient descent step with learning rate $\eta$, the shared expert parameters are updated as:
\begin{equation}
\theta_s^{\prime}
=
\theta_s
-
\eta
\left(
\mathbf{g}_{q}
+
\lambda_{c}\mathbf{g}_{c}
\right).
\end{equation}
Using a first-order approximation, the change in the quality regression objective is
\begin{equation}
\mathcal{L}_{q}(\theta_s^{\prime})
\approx
\mathcal{L}_{q}(\theta_s)
-\eta\|\mathbf{g}_{q}\|_2^2
-\eta\lambda_c
\left\langle
\mathbf{g}_{q},\mathbf{g}_{c}
\right\rangle. 
\end{equation}
The auxiliary labels and continuous quality scores may impose different representation requirements. Therefore, the gradients of the two objectives are not guaranteed to be aligned. The auxiliary update conflicts with quality regression when $\left\langle \mathbf{g}_{q},\mathbf{g}_{c}\right\rangle<0$, since the two gradients then point in opposing directions. Therefore, uniformly applying auxiliary supervision to all shared experts cannot guarantee consistent improvements in quality prediction and may restrict fine-grained quality representation learning.

\noindent\textbf{Hierarchical MoE Design in Life-IQA$^{+}$.}
Motivated by Observation-2 and the above optimization analysis, Life-IQA$^{+}$ restricts the auxiliary classification objective to a subset of shared experts rather than the entire shared expert branch. As shown in Fig.~\ref{fig_3}, the shared expert layer consists of auxiliary shared experts $\mathcal{E}_{\text{aux}}=\{E_{1}^{\text{aux}},E_{2}^{\text{aux}},\ldots,E_{M_a}^{\text{aux}}\}$ and general shared experts $\mathcal{E}_{\text{gen}}=\{E_{1}^{\text{gen}},E_{2}^{\text{gen}},\ldots,E_{M_g}^{\text{gen}}\}$. Both groups contribute to quality regression, while only the auxiliary shared experts receive classification supervision. This separation incorporates auxiliary category information without imposing the potentially conflicting classification objective on every shared expert. The general shared experts remain optimized solely by quality regression, preserving a regression-only path for continuous quality differences.

For synthetic distortion datasets, distortion type labels are used to guide the auxiliary shared experts. For authentic distortion datasets, continuous quality scores are discretized into five quality levels and used as auxiliary labels. Inspired by DeepMoE \coloredcite{eigen2014learningfactoredrepresentationsdeep}, the shared expert layer is followed by a sparse routed expert layer, which performs input-dependent refinement on the combined shared representation without direct auxiliary classification supervision.

All shared experts are densely activated for each quality token $x_i$, and their outputs are aggregated into the shared residual $\Delta_i^s$. The outputs of the auxiliary shared experts are additionally averaged across experts and quality tokens before being passed to a fully connected classifier. The shared update and auxiliary objective are formulated as
\begin{align}
\Delta_i^{s} &=
\frac{1}{M_a}\sum_{m=1}^{M_a} E_{m}^{\text{aux}}(x_i)
+
\frac{1}{M_g}\sum_{n=1}^{M_g} E_{n}^{\text{gen}}(x_i), \\
 x_i^{S} &= x_i+\rho_s \Delta_i^{s},\\
h_{\text{aux}} &=
\frac{1}{N}\sum_{i=1}^{N}
\left(
\frac{1}{M_a}\sum_{m=1}^{M_a} E_{m}^{\text{aux}}(x_i)
\right), \\
z_{\text{cls}} &= W_{\text{cls}}h_{\text{aux}}+b_{\text{cls}}, \\
\mathcal{L}_{\text{cls}} &= \mathrm{CE}(z_{\text{cls}},y_{\text{cls}}),
\end{align}
where $x_i^S$ is the updated representation after the shared expert layer and $\rho_s$ controls the contribution of the shared residual. The pooled representation $h_{\text{aux}}$ is mapped to the classification logits $z_{\text{cls}}$, and $\mathcal{L}_{\text{cls}}$ is computed using the distortion-type or quality level label $y_{\text{cls}}$. 

The sparse routed expert layer consists of a set of routed experts 
$\mathcal{R}=\{R_1,R_2,\cdots,R_{M_r}\}$ and a routing network $G_r(\cdot)$. 
Given the updated quality token $x_i^S$ from the shared expert layer, the router predicts expert-routing logits $a_i=G_r(x_i^S)\in\mathbb{R}^{M_r}$, which measure the affinity between the token and different routed experts. 
To achieve sparse activation, only the Top-$K$ experts are selected for each token, and the routing weights are computed over the selected experts:
\begin{align}
\mathcal{T}_i &= \mathrm{TopK}(a_i,K), \\
g_i^m &=
\begin{cases}
\frac{\exp(a_i^m)}{\sum_{j\in\mathcal{T}_i}\exp(a_i^j)}, & m\in\mathcal{T}_i, \\
0, & m\notin\mathcal{T}_i,
\end{cases}
\end{align}
where $\mathcal{T}_i$ denotes the index set of the selected Top-$K$ experts, and $g_i^m$ is the routing weight assigned to the $m$-th routed expert. 
The routed residual is then obtained by the weighted aggregation of the selected expert outputs:
\begin{equation}
\Delta_i^r=
\sum_{m=1}^{M_r}
g_i^m R_m(x_i^S).
\end{equation}


To preserve the original quality representation and stabilize sparse expert routing, we adopt a residual expert fusion strategy. 
Instead of directly replacing the decoder-level quality token with expert outputs, the expert branches are formulated as incremental residual corrections. 
The shared expert residual provides stable degradation-aware quality priors, while the routed expert residual captures sample-specific quality patterns through sparse expert selection. 
Moreover, since sparse routing may be unstable at the early training stage, explicitly scaling the routed residual helps prevent unreliable routing outputs from dominating the final representation.
The residual expert fusion is formulated as:
\begin{equation}
x_i^{\text{out}}=x_i+\gamma_1\Delta_i^s+\gamma_2\Delta_i^r, 
\end{equation}
where $x_i$ denotes the original quality token, and $x_i^{\text{out}}$ is the output token after hierarchical expert fusion. 
$\Delta_i^s$ and $\Delta_i^r$ represent the shared expert residual and the sparse routed expert residual, respectively. 
$\gamma_1$ and $\gamma_2$ are scaling factors that control the contributions of the two expert branches.

\subsection{Quality Score Regression and Loss Function}
To map the final feature $x^{\text{out}}_{i}$ to the quality score of the image, we feed it into a linear regression layer to obtain the predicted quality score, as illustrated in Fig.~\ref{fig_3}. Each query token produces an individual quality score, and the final image quality score is obtained by averaging the scores of all tokens.


Given a mini-batch of distorted images 
$\mathcal{B}=\{(I_i,y_i,c_i)\}_{i=1}^{B}$, 
where $y_i$ is the ground-truth quality score and $c_i$ denotes the auxiliary class label, the model predicts an image quality score $\hat{y}_i$ and auxiliary classification logits $\mathbf{p}_i$. 
The sparse routed expert layer also produces two router regularization terms, namely the load-balancing loss $\mathcal{L}_{\text{bal}}$ and the router z-loss $\mathcal{L}_{z}$.

\paragraph{Quality Regression Loss}
We use the mean absolute error between the predicted and ground-truth quality scores:
\begin{equation}
\mathcal{L}_{q}
=
\frac{1}{B}
\sum_{i=1}^{B}
\left|
\hat{y}_i-y_i
\right|.
\end{equation}
\paragraph{Auxiliary Classification Loss}
The auxiliary shared experts are trained with an auxiliary classification objective. For the $b$-th image, $c_b$ denotes the distortion type on synthetically distorted datasets or the quality level obtained by discretizing the continuous quality score on authentically distorted datasets. The classification loss is defined as:
\begin{equation}
\mathcal{L}_{\text{cls}}
=
-\frac{1}{B}
\sum_{b=1}^{B}
\log
\frac{
\exp\left(z_b^{\text{cls}}(c_b)\right)
}{
\sum_{k=1}^{C_a}
\exp\left(z_b^{\text{cls}}(k)\right)
},
\end{equation}
where $\mathbf{z}_b^{\text{cls}}\in\mathbb{R}^{C_a}$ denote the logits produced by the auxiliary classification head. $C_a$ denotes the number of auxiliary classes.
\paragraph{Pairwise Ranking Loss}
Optimizing only the absolute regression loss may not explicitly preserve the relative perceptual order among images. 
Therefore, we introduce a batch-wise pairwise ranking loss. 
For each image pair $(I_i,I_j)$ in a mini-batch, the predicted and ground-truth quality differences are defined as:
\begin{equation}
\Delta^{p}_{ij}=\hat{y}_i-\hat{y}_j,
\qquad
\Delta^{g}_{ij}=y_i-y_j. 
\end{equation}
The ranking target is given by:
\begin{equation}
s_{ij}=\mathrm{sign}(\Delta^{g}_{ij}). 
\end{equation}
To avoid ambiguous supervision from image pairs with very similar quality scores, we only retain pairs whose ground-truth quality gap is larger than a threshold $\tau$:
\begin{equation}
m_{ij}
=
\mathbb{I}(|\Delta^{g}_{ij}|>\tau)\cdot \mathbb{I}(i\neq j).
\end{equation}
where $\mathbb{I}(\cdot)$ denotes the indicator function. 
The pairwise ranking loss is formulated as:
\begin{equation}
\mathcal{L}_{\text{rank}}
=
\frac{1}{\sum_{i,j}m_{ij}}
\sum_{i=1}^{B}
\sum_{j=1}^{B}
m_{ij}
\cdot
\mathrm{Softplus}
\left(
-s_{ij}\Delta^{p}_{ij}
\right).
\end{equation}
This loss penalizes inconsistent quality ordering and encourages the predicted scores to preserve the relative perceptual ranking within each mini-batch.

\paragraph{Total Loss}
The overall training objective is formulated as:
\begin{equation}
\mathcal{L}_{\text{total}}
=
\mathcal{L}_{q}
+
\lambda_{\text{cls}}\mathcal{L}_{\text{cls}}
+
\lambda_{\text{rank}}\mathcal{L}_{\text{rank}}
+
\lambda_{\text{bal}}\mathcal{L}_{\text{bal}}
+
\lambda_{z}\mathcal{L}_{z},
\end{equation}
where $\lambda_{\text{cls}}$, $\lambda_{\text{rank}}$, $\lambda_{\text{bal}}$, and
$\lambda_z$ control the contributions of the auxiliary classification,
pairwise ranking, load-balancing, and router z-loss terms, respectively.
\begin{table*}[t]
\centering
\caption{Performance comparison on seven IQA benchmark datasets.
The best results are highlighted in \textcolor{red}{red},
the second-best in \textcolor{blue}{blue},
and the third-best in \textcolor{green}{green}.}
\label{table_1}
\footnotesize 
\setlength{\tabcolsep}{1.8pt}
\renewcommand{\arraystretch}{1.05}
\begin{tabular*}{\textwidth}{@{\extracolsep{\fill}}lc*{14}{c}@{}}
\toprule[0.3mm]
    \multirow{2}{*}{Method} & \multicolumn{1}{c}{Params} &
    \multicolumn{2}{c}{LIVE \coloredcite{1709988}} &
    \multicolumn{2}{c}{CSIQ \coloredcite{larson2010most}} &
    \multicolumn{2}{c}{TID2013 \coloredcite{PONOMARENKO201557}}  &
    \multicolumn{2}{c}{KADID-10K \coloredcite{8743252}}  &
    \multicolumn{2}{c}{LIVEC \coloredcite{7327186}}  &
    \multicolumn{2}{c}{KonIQ-10K \coloredcite{8968750}}  &
    \multicolumn{2}{c}{SPAQ \coloredcite{9156490}}  \\
    \cmidrule(lr){3-4}
    \cmidrule(lr){5-6}
    \cmidrule(lr){7-8}
    \cmidrule(lr){9-10}
    \cmidrule(lr){11-12}
    \cmidrule(lr){13-14}
    \cmidrule(lr){15-16}
    \multicolumn{1}{c}{} & \multicolumn{1}{c}{M} &
    \multicolumn{1}{c}{SRCC} & \multicolumn{1}{c}{PLCC} &
    \multicolumn{1}{c}{SRCC} & \multicolumn{1}{c}{PLCC} &
    \multicolumn{1}{c}{SRCC} & \multicolumn{1}{c}{PLCC} &
    \multicolumn{1}{c}{SRCC} & \multicolumn{1}{c}{PLCC} &
    \multicolumn{1}{c}{SRCC} & \multicolumn{1}{c}{PLCC} &
    \multicolumn{1}{c}{SRCC} & \multicolumn{1}{c}{PLCC} &
    \multicolumn{1}{c}{SRCC} & \multicolumn{1}{c}{PLCC} \\
    \midrule
    MUSIQ \coloredcite{9710973} & 27 &
      0.940 & 0.911 & 0.871 & 0.893 & 0.773 & 0.815 & 0.875 & 0.872 & 0.702 & 0.746 & 0.916 & 0.928 & 0.918 & 0.921 \\
    TRes \coloredcite{9706735} & 152 &
      0.969 & 0.968 & 0.922 & 0.942 & 0.863 & 0.883 & 0.859 & 0.858 & 0.846 & 0.877 & 0.915 & 0.928 & - & - \\
    DACNN \coloredcite{9817377} & - &
      0.978 & 0.980 & 0.943 & 0.957 & 0.871 & 0.889 & 0.905 & 0.905 & 0.866 & 0.884 & 0.901 & 0.912 & 0.915 & 0.921 \\
    DEIQT \coloredcite{qin2023dataefficientimagequalityassessment} & 24 &
      0.980 & \textbf{\textcolor{green}{0.982}} & 0.946 & \textbf{\textcolor{green}{0.963}} & 0.892 & 0.908 & 0.889 & 0.887 & 0.875 & 0.894 & 0.921 & 0.934 & 0.919 & 0.923 \\
    CICI \coloredcite{10213444} & - &
      0.977 & 0.979 & 0.930 & 0.951 & 0.865 & 0.874 & - & - & 0.865 & 0.874 & 0.921 & 0.930 & - & - \\
    PCIQA \coloredcite{10520932} & - &
      0.976 & 0.976 & 0.931 & 0.942 & - & - & 0.870 & 0.866 & 0.863 & 0.880 & 0.918 & 0.930 & - & - \\
    TempQT \coloredcite{10287911} & - &
      0.976 & 0.977 & 0.950 & 0.960 & 0.883 & 0.906 & - & - & 0.870 & 0.886 & 0.915 & 0.928 & - & - \\
    FsPN \coloredcite{10706861} & 34 &
      0.979 & 0.979 & 0.955 & 0.962 & 0.897 & 0.912 & 0.911 & 0.909 & 0.860 & 0.884 & 0.918 & 0.932 & - & - \\
    LIQE \coloredcite{10204315} & 151 &
      0.970 & 0.951 & 0.936 & 0.939 & - & - & 0.930 & 0.931 & \textbf{\textcolor{red}{0.904}} & \textbf{\textcolor{green}{0.910}} & 0.919 & 0.908 & - & - \\
    Re-IQA \coloredcite{10204004} & 48 &
      0.970 & 0.971 & 0.947 & 0.960 & 0.880 & 0.901 & 0.908 & 0.912 & - & - & - & - & 0.918 & \textbf{\textcolor{green}{0.925}} \\
    QCN \coloredcite{10656546} & - &
      - & - & - & - & - & - & - & - & 0.875 & 0.893 & \textbf{\textcolor{blue}{0.934}} & \textbf{\textcolor{blue}{0.945}} & \textbf{\textcolor{green}{0.923}} & \textbf{\textcolor{blue}{0.928}} \\
    LoDa \coloredcite{10655501} & 118 &
      0.975 & 0.979 & - & - & 0.869 & 0.901 & 0.931 & 0.936 & 0.876 & 0.899 & \textbf{\textcolor{green}{0.932}} & \textbf{\textcolor{green}{0.944}} & \textbf{\textcolor{blue}{0.925}} & \textbf{\textcolor{blue}{0.928}} \\
    VISGA \coloredcite{10845143} & - &
      \textbf{\textcolor{red}{0.987}} & \textbf{\textcolor{red}{0.988}} &
      \textbf{\textcolor{green}{0.960}} & \textbf{\textcolor{blue}{0.971}} &
      \textbf{\textcolor{green}{0.901}} & \textbf{\textcolor{green}{0.914}} &
      \textbf{\textcolor{green}{0.939}} & \textbf{\textcolor{blue}{0.944}} &
      0.882 & \textbf{\textcolor{blue}{0.912}} &
      0.931 & 0.940 & - & - \\
    \midrule[0.3mm]
    {Life-IQA} \coloredcite{Tang_2026_CVPR} & 95 &
      \textbf{\textcolor{green}{0.982}} & \textbf{\textcolor{green}{0.982}} &
      \textbf{\textcolor{blue}{0.966}} & \textbf{\textcolor{blue}{0.971}} &
      \textbf{\textcolor{blue}{0.918}} & \textbf{\textcolor{blue}{0.930}} &
      \textbf{\textcolor{blue}{0.940}} & \textbf{\textcolor{green}{0.943}} &
      \textbf{\textcolor{blue}{0.896}} & \textbf{\textcolor{red}{0.919}} &
      \textbf{\textcolor{red}{0.937}} & \textbf{\textcolor{red}{0.948}} &
      \textbf{\textcolor{red}{0.926}} & \textbf{\textcolor{red}{0.929}} \\
    
    {Life-IQA$^{+}$} & 45 &
      \textbf{\textcolor{blue}{0.983}} & \textbf{\textcolor{blue}{0.984}} &
      \textbf{\textcolor{red}{0.970}} & \textbf{\textcolor{red}{0.973}} &
      \textbf{\textcolor{red}{0.941}} & \textbf{\textcolor{red}{0.948}} &
      \textbf{\textcolor{red}{0.945}} & \textbf{\textcolor{red}{0.948}} &
      \textcolor{green}{0.893} & \textbf{\textcolor{blue}{0.912}} &
      \textbf{\textcolor{red}{0.937}} & \textbf{\textcolor{blue}{0.945}} &
      \textbf{\textcolor{blue}{0.925}} & \textbf{\textcolor{red}{0.929}} \\
    \bottomrule[0.3mm]
\end{tabular*}

\end{table*}


\begin{table*}[htbp]
\centering
\caption{Cross-dataset evaluation of BIQA methods under different train-test dataset settings.}
\label{table_2}
\footnotesize 
\setlength{\tabcolsep}{2.0pt}
\renewcommand{\arraystretch}{1.05}
\begin{tabular*}{0.85\textwidth}{@{\extracolsep{\fill}}l*{12}{c}@{}}
\toprule[0.3mm]
    \textit{Train on:} &
    \multicolumn{2}{c}{CSIQ \coloredcite{larson2010most}} &
    \multicolumn{6}{c}{KADID-10K \coloredcite{8743252}}  &
    \multicolumn{2}{c}{LIVEC \coloredcite{7327186}}  &
    \multicolumn{2}{c}{SPAQ \coloredcite{9156490}}  \\
    \cmidrule(lr){2-3}
    \cmidrule(lr){4-9}
    \cmidrule(lr){10-11}
    \cmidrule(lr){12-13}
    \textit{Test on:} &
    \multicolumn{2}{c}{LIVE \coloredcite{1709988}}  &
    \multicolumn{2}{c}{CSIQ \coloredcite{larson2010most}} &
    \multicolumn{2}{c}{TID2013 \coloredcite{PONOMARENKO201557}} &
    \multicolumn{2}{c}{LIVE \coloredcite{1709988}} &
    \multicolumn{2}{c}{SPAQ \coloredcite{9156490}} &
    \multicolumn{2}{c}{LIVEC \coloredcite{7327186}} \\
    \cmidrule(lr){2-3}
    \cmidrule(lr){4-5}
    \cmidrule(lr){6-7}
    \cmidrule(lr){8-9}
    \cmidrule(lr){10-11}
    \cmidrule(lr){12-13}
    {Method / Metric} &
    \multicolumn{1}{c}{SRCC} & \multicolumn{1}{c}{PLCC} &
    \multicolumn{1}{c}{SRCC} & \multicolumn{1}{c}{PLCC} &
    \multicolumn{1}{c}{SRCC} & \multicolumn{1}{c}{PLCC} &
    \multicolumn{1}{c}{SRCC} & \multicolumn{1}{c}{PLCC} &
    \multicolumn{1}{c}{SRCC} & \multicolumn{1}{c}{PLCC} &
    \multicolumn{1}{c}{SRCC} & \multicolumn{1}{c}{PLCC} \\
    \midrule
    HyperIQA \coloredcite{9156687} &
    0.941 & 0.932 &
    0.795 & 0.818 &
    0.708 & 0.721 &
    0.921 & 0.902 &
    0.840 & 0.838 &
    0.755 & 0.794 \\
    DACNN \coloredcite{9817377} &
    0.940 & 0.938 &
    0.825 & 0.833 &
    0.669 & 0.690 &
    0.898 & 0.874 &
    0.854 & 0.852 &
    0.799 & 0.808 \\
    FsPN \coloredcite{10706861} &
    0.948 & 0.945 &
    0.814 & 0.831 &
    0.686 & 0.707 &
    0.925 & 0.912 &
    0.833 & 0.827 &
    0.702 & 0.728 \\
    DEIQT \coloredcite{qin2023dataefficientimagequalityassessment} &
    0.943 & 0.930 &
    0.795 & 0.820 &
    0.713 & 0.732 &
    0.932 & 0.908 &
    0.854 & 0.852 &
    0.781 & 0.803 \\
    Life-IQA \coloredcite{Tang_2026_CVPR} &
    0.946 & 0.935 &
    0.792 & 0.824 &
    \textbf{\textcolor[rgb]{.933,0,0}{0.767}} &
    \textbf{\textcolor[rgb]{.933,0,0}{0.774}} &
    0.936 & 0.915 &
    0.853 & 0.855 &
    0.944 & 0.937 \\
    Life-IQA$^{+}$ &
    \textbf{\textcolor[rgb]{.933,0,0}{0.953}} &
    \textbf{\textcolor[rgb]{.933,0,0}{0.950}} &
    \textbf{\textcolor[rgb]{.933,0,0}{0.843}} &
    \textbf{\textcolor[rgb]{.933,0,0}{0.860}} &
    0.736 & 0.735 &
    \textbf{\textcolor[rgb]{.933,0,0}{0.944}} &
    \textbf{\textcolor[rgb]{.933,0,0}{0.928}} &
    \textbf{\textcolor[rgb]{.933,0,0}{0.856}} &
    \textbf{\textcolor[rgb]{.933,0,0}{0.856}} &
    \textbf{\textcolor[rgb]{.933,0,0}{0.945}} &
    \textbf{\textcolor[rgb]{.933,0,0}{0.950}} \\
    \bottomrule[0.3mm]
\end{tabular*}
\vspace{-5.80pt}
\end{table*}

\section{Experiments}
\subsection{Experimental Setting}
We evaluate our method on seven publicly available datasets. These include four synthetic distortion datasets: LIVE \coloredcite{1709988}, CSIQ \coloredcite{larson2010most}, KADID-10K \coloredcite{8743252} and TID2013 \coloredcite{PONOMARENKO201557} and three authentic distortion datasets including LIVEC \coloredcite{7327186}, SPAQ \coloredcite{9156490} and KonIQ-10K \coloredcite{8968750}. Specifically, the LIVE dataset contains 779 distorted images; CSIQ includes 866 images with 6 distortion types; TID2013 comprises 3,000 images involving 24 distortion types; and KADID-10K consists of 10,125 images corresponding to 25 distortion types. In addition, LIVEC contains 1,162 images, KonIQ-10K includes 10,073 images, and SPAQ comprises 11,125 images. The learning rate and number of training epochs used for each dataset are specified below. Specifically, the learning rate is set to $1.0\times10^{-4}$ for LIVE, CSIQ, and TID2013, $8.0\times10^{-5}$ for KADID-10K, and $2.0\times10^{-5}$ for authentic distortion datasets. Unless otherwise specified, the Transformer decoder comprises $4$ layers, the query set size is $6$, the shared expert layer uses $4$ auxiliary shared experts and $4$ general shared experts, the sparse routed expert layer uses $8$ experts with Top-$K{=}2$, and the decoder adopts multi-head attention with $6$ heads and an embedding dimension of $384$. The model is trained for 25 to 80 epochs depending on the dataset scale. The batch size is set to 32 for small synthetic datasets, 64 for LIVEC, and 128 for large-scale datasets.

Like other IQA methods, we evaluate our test results using Pearson linear correlation coefficient (PLCC) and Spearman rank correlation coefficient (SRCC). Higher values for both metrics indicate better performance. We compare Life-IQA$^{+}$ with state-of-the-art (SOTA) methods, including DBCNN \coloredcite{8576582}, MMMNet \coloredcite{9337868}, VCRNet \coloredcite{9694502}, MUSIQ \coloredcite{9710973}, TRes \coloredcite{9706735}, DACNN \coloredcite{9817377}, DEIQT \coloredcite{qin2023dataefficientimagequalityassessment}, CICI \coloredcite{10213444}, PCIQA \coloredcite{10520932}, TempQT \coloredcite{10287911}, FsPN \coloredcite{10706861}, LIQE \coloredcite{10204315}, Re-IQA \coloredcite{10204004}, QCN \coloredcite{10656546}, LoDa \coloredcite{10655501}, VISGA \coloredcite{10845143}, Life-IQA \coloredcite{Tang_2026_CVPR}. The input distorted images were randomly cropped into small segments measuring $224 \times 224$ pixels. We randomly split each dataset into $0.8$ training set and $0.2$ testing set, and repeated the experiments $10$ times, reporting the median values of SRCC and PLCC. 







\subsection{Comparisons with the State-of-the-art Methods}

Table~\ref{table_1} compares Life-IQA$^{+}$ with representative SOTA BIQA methods on seven benchmark datasets. Life-IQA$^{+}$ employs Swin-Tiny as its backbone, initialized with weights pretrained on ImageNet-22K and fine-tuned on ImageNet-1K, and contains 45M parameters. The main observations are summarized as follows.
On the synthetically distorted datasets, Life-IQA$^{+}$ achieves the best SRCC results on CSIQ (0.970), TID2013 (0.941), and KADID-10K (0.945), while ranking second on LIVE (0.983). These results demonstrate consistently high prediction accuracy across diverse synthetic distortion types and levels.
On the authentically distorted datasets, Life-IQA$^{+}$ obtains SRCC values of 0.893 on LIVEC, 0.937 on KonIQ-10K, and 0.925 on SPAQ. The proposed model achieves the tied-best SRCC on KonIQ-10K. Life-IQA$^{+}$ also matches or outperforms QCN, LoDa, and VISGA on SRCC across these datasets. The results confirm its consistently competitive performance under both synthetic and authentic distortion settings.
Compared with Life-IQA, Life-IQA$^{+}$ reduces the parameter count from 95M to 45M while improving the average SRCC across the seven datasets from 0.938 to 0.942. The improvement is particularly evident on TID2013, with an SRCC gain of 0.023. Life-IQA$^{+}$ therefore provides a more favorable balance between prediction accuracy and model complexity.

\begin{figure*}[t]
\label{fig_6}
\centering
\includegraphics[width=7in]{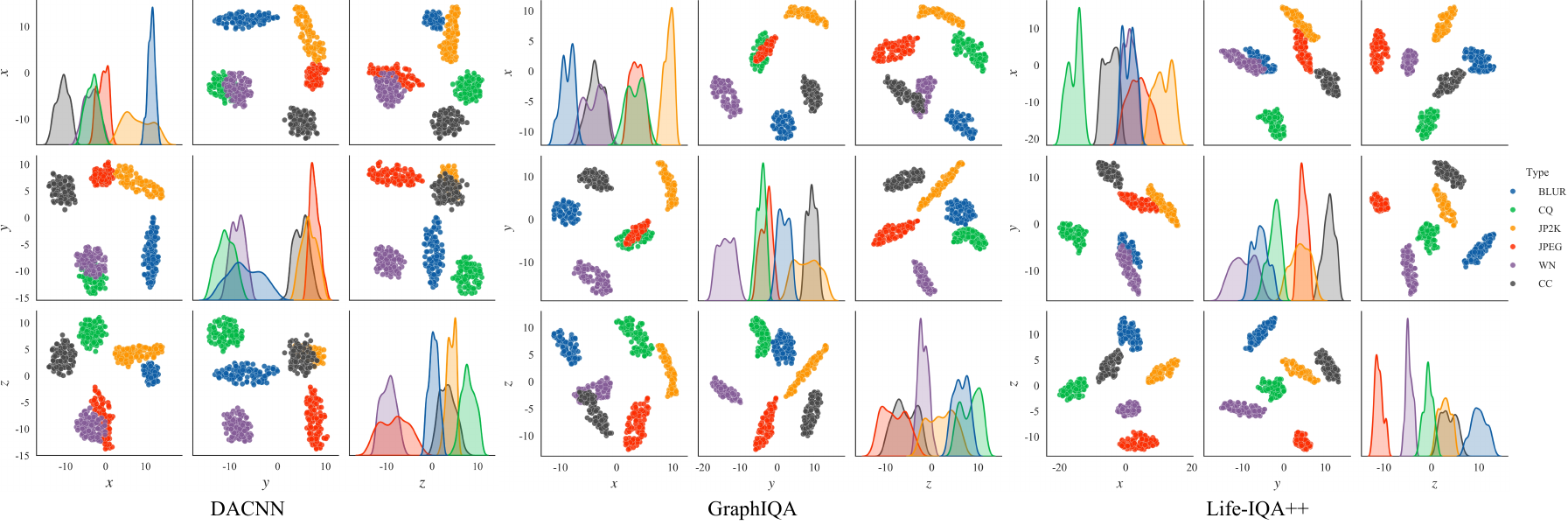}
\caption{Comparison of feature representations learned by DACNN \coloredcite{9817377}, GraphIQA \coloredcite{9720138}, and Life-IQA$^{+}$ for six common distortion types. Life-IQA$^{+}$ produces more compact within-type distributions and clearer separation between distortion types.}
\vspace{-3mm}
\label{fig_6}
\end{figure*}

\begin{figure}[t]
\label{sec:intro}
\centering
\includegraphics[width= 3.5in]{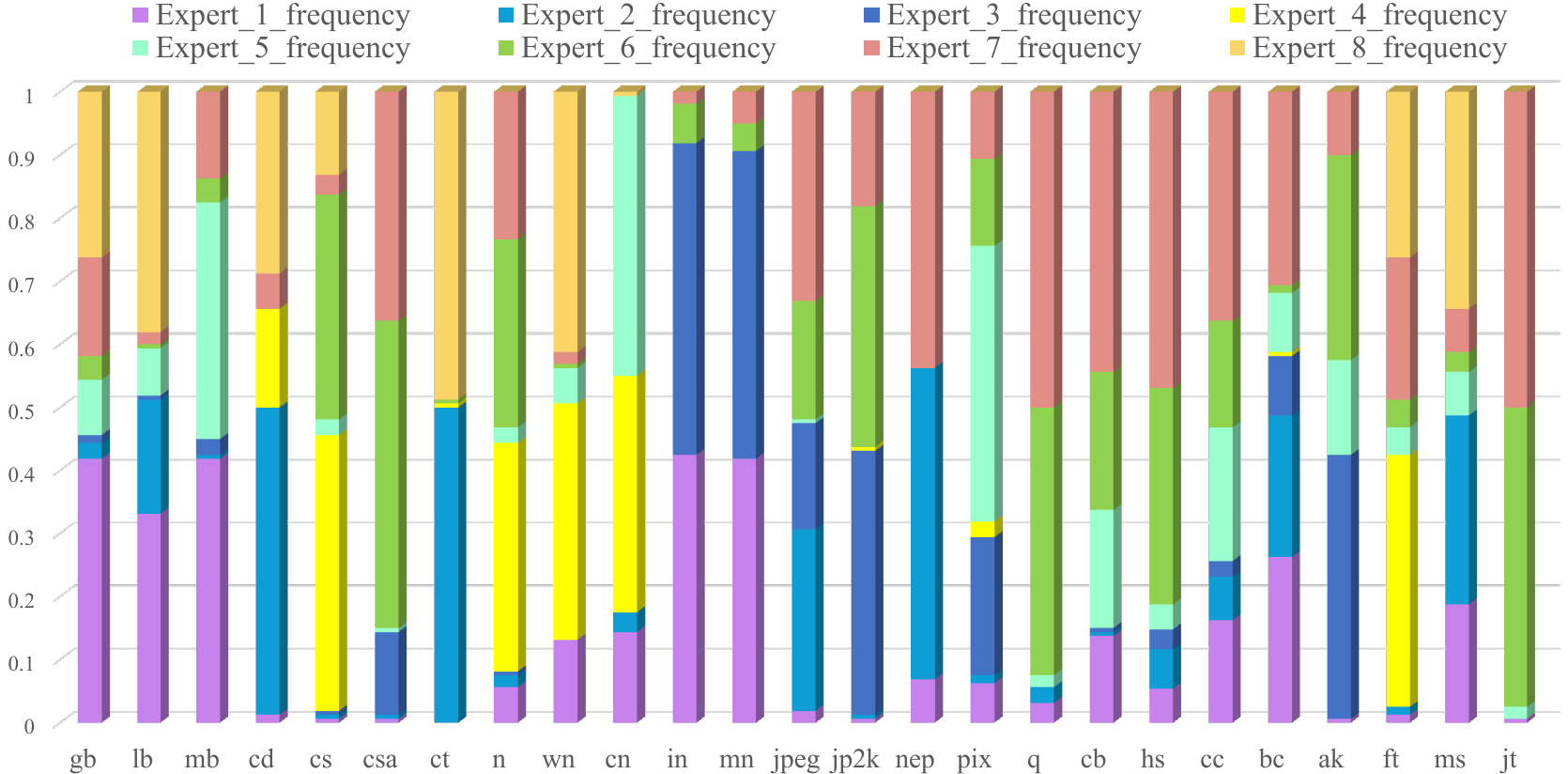}
\caption{Normalized Top-$K$ expert-selection frequencies for 25 distortion types on KADID-10K. Each bar aggregates the routing decisions over all samples of one distortion category.}
\vspace{-5mm}
\label{fig_7}
\end{figure}


\subsection{Cross-Dataset Evaluation}
We further assess cross-dataset generalization by training each model on a source dataset and directly evaluating it on a different target dataset, without using target-domain samples for training or model adaptation. As reported in Table~\ref{table_2}, Life-IQA$^{+}$ achieves the highest SRCC in five of the six transfer settings. For synthetic-to-synthetic transfer, Life-IQA$^{+}$ improves SRCC over Life-IQA from 0.792 to 0.843 on CSIQ and from 0.936 to 0.944 on LIVE when trained on KADID-10K. With CSIQ as the training set and LIVE as the test set, SRCC increases from 0.946 to 0.953. For authentic-to-authentic transfer, Life-IQA$^{+}$ improves SRCC from 0.853 to 0.856 when trained on LIVEC and tested on SPAQ, and from 0.944 to 0.945 in the reverse setting. The only exception occurs when training on KADID-10K and testing on TID2013, where Life-IQA$^{+}$ achieves an SRCC of 0.736 compared 
\begin{table}[htbp]
\centering
\caption{Data-efficient learning validation with the training set
containing 20\%, 40\% and 60\% images.}
\label{table_4}
\footnotesize 
\setlength{\tabcolsep}{2.5pt}
\renewcommand{\arraystretch}{1.4}
\begin{tabular*}{\columnwidth}{@{\extracolsep{\fill}}cl*{6}{c}@{}}
\toprule
    \multicolumn{2}{c}{} & \multicolumn{2}{c}{KonIQ-10K} & \multicolumn{2}{c}{LIVEC} & \multicolumn{2}{c}{LIVE} \\
\cmidrule{3-8}    \multicolumn{1}{c}{Mode} & \multicolumn{1}{c}{Methods} & \multicolumn{1}{c}{SRCC} & \multicolumn{1}{c}{PLCC} & \multicolumn{1}{c}{SRCC} & \multicolumn{1}{c}{PLCC} & \multicolumn{1}{c}{SRCC} & \multicolumn{1}{c}{PLCC} \\
    \midrule
    \multirow{4}{*}{20\%} & HyperIQA \coloredcite{9156687} & 0.869 & 0.873 & 0.776 & 0.809 & 0.951 & 0.950 \\
          &  DEIQT \coloredcite{qin2023dataefficientimagequalityassessment} & 0.888 & 0.908 & 0.792 & 0.822 & 0.965 & 0.968 \\
          & Life-IQA \coloredcite{Tang_2026_CVPR}  &  0.908     & \textbf{\textcolor{red}{0.923}}      & 0.804 & 0.842 &  \textbf{\textcolor{red}{0.970}}    & 0.970 \\
          & Life-IQA$^{+}$  &  \textbf{\textcolor{red}{0.910}}     & 0.922      & \textbf{\textcolor{red}{0.816}} & \textbf{\textcolor{red}{0.851}} &  \textbf{\textcolor{red}{0.970}}     & \textbf{\textcolor{red}{0.972}} \\
    \midrule
    \multirow{4}{*}{40\%} & HyperIQA \coloredcite{9156687} & 0.892 & 0.908 & 0.832 & 0.849 & 0.959 & 0.961 \\
          &  DEIQT \coloredcite{qin2023dataefficientimagequalityassessment} & 0.903 & 0.922 & 0.838 & 0.855 & 0.971 & 0.973 \\
          & Life-IQA \coloredcite{Tang_2026_CVPR}  &   0.920    &   0.930    &    \textbf{\textcolor{red}{0.855}}    &   0.880  &  0.971    &  0.972 \\
          & Life-IQA$^{+}$     &\textbf{\textcolor{red}{0.923}} &  \textbf{\textcolor{red}{0.935}}       & 0.844 &  \textbf{\textcolor{red}{0.882}} &  \textbf{\textcolor{red}{0.975}}     & \textbf{\textcolor{red}{0.976}} \\
    \midrule
    \multirow{4}{*}{60\%} & HyperIQA \coloredcite{9156687} & 0.901 & 0.914 & 0.843 & 0.862 & 0.960  & 0.960 \\
          &  DEIQT \coloredcite{qin2023dataefficientimagequalityassessment} & 0.914 & 0.931 & 0.848 & 0.877 & 0.972 & 0.974 \\
          & Life-IQA \coloredcite{Tang_2026_CVPR}  &   0.930   &  0.939    &\textbf{\textcolor{red}{0.882}}     &\textbf{\textcolor{red}{0.898}}      &  0.973    & 0.974  \\
          & Life-IQA$^{+}$  &   \textbf{\textcolor{red}{0.931}}    &   \textbf{\textcolor{red}{0.941}}    &0.867  &0.887  &\textbf{\textcolor{red}{0.981}}       & \textbf{\textcolor{red}{0.982}} \\
    \bottomrule
\end{tabular*}
\vspace{-2mm}
\end{table}

with 0.767 for Life-IQA, although it still ranks second among the compared methods. The largest improvement is observed when transferring from KADID-10K to CSIQ, with an absolute SRCC gain of 0.051. These results suggest improved cross-dataset generalization in most of the evaluated settings.

\begin{table}[t]
\centering
\caption{Comparison of model complexity and computational efficiency.}
\label{table_3}
\footnotesize 
\setlength{\tabcolsep}{3.0pt}
\renewcommand{\arraystretch}{1.0}
\begin{tabular*}{\columnwidth}{@{\extracolsep{\fill}}lcccc@{}}
\toprule
    Method & Params & Train-FPS & Test-FPS & CSIQ \\
    \midrule
    Life-IQA   &95M  &181.6  &617.6  &0.966  \\
    Life-IQA$^{+}$ &45M  &298.9  &1149.8  &0.970  \\
    \bottomrule
\end{tabular*}
\vspace{-5mm}
\end{table}

\subsection{Representation Visualization and Expert Routing Analysis}

\begin{table}[t]
\centering
\caption{Cross-layer pairing study: deep and shallow interactions. }
\label{table_8}
\footnotesize 
\setlength{\tabcolsep}{2.5pt}
\renewcommand{\arraystretch}{1.4}
\begin{tabular*}{\columnwidth}{@{\extracolsep{\fill}}l*{6}{c}@{}}
\toprule
    \multirow{2}[4]{*}{Method} 
    & \multicolumn{2}{c}{CSIQ} 
    & \multicolumn{2}{c}{LIVE} 
    & \multicolumn{2}{c}{LIVEC} \\
\cmidrule{2-7}    
    \multicolumn{1}{c}{} 
    & \multicolumn{1}{c}{SRCC} 
    & \multicolumn{1}{c}{PLCC} 
    & \multicolumn{1}{c}{SRCC} 
    & \multicolumn{1}{c}{PLCC} 
    & \multicolumn{1}{c}{SRCC} 
    & \multicolumn{1}{c}{PLCC} \\
    \midrule
    $S2-S1$ & 0.950 &0.958  &0.976  & 0.980 &0.760  & 0.792 \\

    $S3-S1$ &0.963  & 0.968 & 0.974 & 0.975 &0.873  &0.894  \\
    
    $S3-S2$ &0.963  & 0.971 &0.980  & 0.981 &0.880  & 0.896 \\
   
    $S4-S1$ & 0.964 & 0.971 &0.980  & 0.983 &0.881  &0.906  \\

    $S4-S2$ & 0.965 & 0.972 & 0.981 & 0.983 &0.883  & 0.903 \\

    $S4-(S1,S2,S3)$ &0.965  &0.972  &0.980  &0.983  & 0.880 & 0.900 \\
    
    Life-IQA$^{+}$ & \textbf{\textcolor[rgb]{.933,0,0}{0.970}} & \textbf{\textcolor[rgb]{.933,0,0}{0.973}} & \textbf{\textcolor[rgb]{.933,0,0}{0.983}} & \textbf{\textcolor[rgb]{.933,0,0}{0.984}} & \textbf{\textcolor[rgb]{.933,0,0}{0.893}} & \textbf{\textcolor[rgb]{.933,0,0}{0.912}} \\
    \bottomrule
\end{tabular*}
\vspace{-3mm}
\end{table}

Fig.~\ref{fig_6} compares low-dimensional projections of the representations learned by DACNN, GraphIQA, and Life-IQA$^{+}$ for six common distortion types. The scatter plots show the distributions of different distortion types across pairs of projected feature dimensions, while the density curves show their distributions along each individual dimension. DACNN produces several dispersed or elongated groups, whereas GraphIQA forms more organized distributions but retains noticeable overlap between some distortion types. Life-IQA$^{+}$ generally yields more compact within-type distributions and clearer separation between types. These results provide qualitative evidence that Life-IQA$^{+}$ learns a more structured distortion-aware representation. To examine the specialization of the routed experts, Fig.~\ref{fig_7} reports the normalized Top-$K$ expert selection frequencies for 25 distortion types on KADID-10K.
Each bar aggregates the routing decisions for all samples of one distortion type and is normalized to sum to one. Clear routing differences can be observed across categories. Specifically, samples with Gaussian blur (GB) and lens blur (LB) are routed mainly to Experts 1 and 8. JPEG distributes its routing load across Experts 2, 3, 6, and 7, whereas JP2K places larger proportions on Experts 3 and 6. These results show that different distortions use different expert combinations. Meanwhile, related distortion types reuse some experts with different proportions, indicating differentiated expert specialization rather than a fixed one-to-one mapping between distortion types and experts. Together, the two visualizations show that Life-IQA$^{+}$ learns distortion discriminative representations while maintaining differentiated expert utilization in the sparse routed layer.

\subsection{Data and Computational Efficiency}

We evaluate data efficiency by training each method with 20\%, 40\%, or 60\% of the original training images on KonIQ-10K, LIVEC, and LIVE. As reported in Table~\ref{table_4}, with only 20\% of the training data, Life-IQA$^{+}$ achieves SRCC values of 0.910 on KonIQ-10K, 0.816 on LIVEC, and 0.970 on LIVE. At both the 40\% and 60\% ratios, Life-IQA$^{+}$ obtains the best SRCC on KonIQ-10K and LIVE. On LIVEC, Life-IQA$^{+}$ achieves the highest SRCC at a training ratio of 20\%, while Life-IQA performs better at 40\% and 60\%. These results show that Life-IQA$^{+}$ remains competitive when trained with limited annotations.

Table~\ref{table_3} compares model complexity and processing speed. Training and inference speeds (FPS) are measured with a batch size of 32 on a single NVIDIA GeForce RTX 3090 GPU. Compared with Life-IQA, Life-IQA$^{+}$ reduces the parameter count from 95M to 45M, while increasing the training throughput from 181.6 to 298.9 FPS and the inference throughput from 617.6 to 1149.8 FPS. Meanwhile, the SRCC on CSIQ increases from 0.966 to 0.970. Life-IQA$^{+}$ provides a better balance between prediction accuracy and computational cost.

\begin{table}[t]
\centering
\caption{Ablation of auxiliary-supervision placement and decoder configuration.}
\label{table_9}
\scriptsize
\setlength{\tabcolsep}{3pt}
\renewcommand{\arraystretch}{1.57}
\begin{tabular}{lcccccc}
\toprule
\multirow{2}{*}{Variant} 
& \multicolumn{2}{c}{LIVE} 
& \multicolumn{2}{c}{TID2013} 
& \multicolumn{2}{c}{KonIQ-10K} \\
\cmidrule(lr){2-3}
\cmidrule(lr){4-5}
\cmidrule(lr){6-7}
& SRCC & PLCC 
& SRCC & PLCC 
& SRCC & PLCC \\
\midrule
No Auxiliary Supervision                  & 0.978     & 0.979     &0.921      & 0.927     & 0.934     &0.943     \\
Layer Interaction Supervision & 0.975     &  0.977    & 0.928     &0.936      & 0.935     &0.944      \\
Hierarchical MoE Supervision  &  0.982    & 0.984     & 0.921     & 0.933     &  0.936    & 0.944     \\
All Shared Experts Supervision                 &0.976      & 0.978     & 0.907     & 0.924     & 0.934     & 0.942     \\
Transformer Decoder Supervision & 0.980     & 0.981     & 0.889     & 0.909     &  0.904    & 0.920     \\
Life-IQA$^{+}$         & \textbf{\textcolor[rgb]{.933,0,0}{0.983}}     & \textbf{\textcolor[rgb]{.933,0,0}{0.984}}     & \textbf{\textcolor[rgb]{.933,0,0}{0.941}}     & \textbf{\textcolor[rgb]{.933,0,0}{0.948}}     &\textbf{\textcolor[rgb]{.933,0,0}{0.937}}      & \textbf{\textcolor[rgb]{.933,0,0}{0.945}}     \\
\bottomrule
\end{tabular}
\vspace{-3mm}
\end{table}

\begin{table}[t]
\centering
\caption{Effects of the numbers of experts and layers on the performance of Life-IQA$^{+}$.}
\label{table_10}
\footnotesize 
\setlength{\tabcolsep}{2.5pt}
\renewcommand{\arraystretch}{1.4}
\begin{tabular*}{\columnwidth}{@{\extracolsep{\fill}}ll*{6}{c}@{}}
\toprule[0.3mm]
\multicolumn{2}{l}{Test on}
& \multicolumn{2}{c}{CSIQ} & \multicolumn{2}{c}{LIVE} & \multicolumn{2}{c}{LIVEC} \\
\cmidrule(lr){3-4}\cmidrule(lr){5-6}\cmidrule(lr){7-8}
\multicolumn{2}{l}{Metric} & SRCC & PLCC & SRCC & PLCC & SRCC & PLCC \\
\midrule
\multirow{3}{*}{Num. Experts}
& 4 & 0.959 & 0.967 & 0.978 & 0.980 & 0.883 & 0.902 \\
& 8 & \textbf{\textcolor{red}{0.970}} & \textbf{\textcolor{red}{0.973}} & \textbf{\textcolor{red}{0.983}} & \textbf{\textcolor{red}{0.984}} & \textbf{\textcolor{red}{0.893}} & \textbf{\textcolor{red}{0.912}} \\
& 12 & 0.961 & 0.970 & 0.982 & 0.983 & 0.889 & 0.908 \\
\midrule
\multirow{3}{*}{Num. Layers}
& 2 & 0.965 & 0.969 & 0.980 & 0.981 & 0.889 & 0.908 \\
& 4 & \textbf{\textcolor{red}{0.970}} & \textbf{\textcolor{red}{0.973}} & \textbf{\textcolor{red}{0.983}} & \textbf{\textcolor{red}{0.984}} & 0.893 & 0.912 \\
& 8 & 0.968 & 0.973 & 0.981 & 0.982 & \textbf{\textcolor{red}{0.896}} & \textbf{\textcolor{red}{0.916}} \\
\bottomrule[0.3mm]
\end{tabular*}

\end{table}

\subsection{Multi-stage feature interaction Analysis}
To examine the choice of interacting backbone stages, we replace the original $S_4\!-\!S_3$ interaction in Life-IQA$^{+}$ with alternative stage combinations while keeping the remaining components unchanged. Here, $S_1$ through $S_4$ denote the backbone stages from shallow to deep. The notation $S_i\!-\!S_j$ represents the interaction between two stages, while $S_4\!-\!(S_1,S_2,S_3)$ first aggregates $S_1$, $S_2$, and $S_3$ before interacting with $S_4$.
As reported in Table~\ref{table_8}, the original $S_4\!-\!S_3$ interaction used by Life-IQA$^{+}$ achieves the best SRCC on all three datasets, reaching 0.970 on CSIQ, 0.983 on LIVE, and 0.893 on LIVEC. In contrast, the shallowest combination, $S_2\!-\!S_1$, obtains only 0.950 on CSIQ and 0.760 on LIVEC. Introducing deeper stages substantially improves the results, particularly on LIVEC. Moreover, aggregating all preceding stages does not provide additional gains: $S_4\!-\!(S_1,S_2,S_3)$ matches $S_4\!-\!S_2$ on CSIQ and performs slightly worse on LIVE and LIVEC. These results verify that interaction between the two deepest backbone stages is more effective than the tested shallow-stage and broader multi-stage alternatives.
\begin{table*}[htbp]
\centering
\caption{Comparison of decoder variants on four IQA benchmark datasets.}
\label{table_6}
\footnotesize 
\setlength{\tabcolsep}{4.0pt}
\renewcommand{\arraystretch}{1.05}
\begin{tabular*}{0.78\textwidth}{@{\extracolsep{\fill}}l*{8}{c}@{}}
\toprule
    \multicolumn{1}{c}{Test on} &
    \multicolumn{2}{c}{LIVE} &
    \multicolumn{2}{c}{TID2013} &
    \multicolumn{2}{c}{KonIQ-10K} &
    \multicolumn{2}{c}{SPAQ} \\
    \cmidrule(lr){2-3}
    \cmidrule(lr){4-5}
    \cmidrule(lr){6-7}
    \cmidrule(lr){8-9}
    \multicolumn{1}{c}{Method} &
    \multicolumn{1}{c}{SRCC} & \multicolumn{1}{c}{PLCC} &
    \multicolumn{1}{c}{SRCC} & \multicolumn{1}{c}{PLCC} &
    \multicolumn{1}{c}{SRCC} & \multicolumn{1}{c}{PLCC} &
    \multicolumn{1}{c}{SRCC} & \multicolumn{1}{c}{PLCC} \\
    \midrule
        SwinT +  Decoder &0.958  &0.957  
     &0.844  &0.873
     &0.860  &0.882
     &0.894  &0.894\\
     SwinT + Parallel MoE 
     &0.980  &0.980 
     &0.924  &0.934
     &0.933  &0.941
     &0.924  &0.928\\
    GCN$\rightarrow$ MHA  & 0.981&0.981
     & 0.930 &0.933
     & 0.933 &0.943
     &0.922  &0.926\\
    MoE$\rightarrow$ FFN 
     & 0.979 &0.979
     & 0.925&0.932
     &0.930  &0.940
     & 0.920 &0.925\\
    \midrule
    Life-IQA$^{+}$ &
    \textbf{\textcolor[rgb]{.933,0,0}{0.983}} &
    \textbf{\textcolor[rgb]{.933,0,0}{0.984}} &
    \textbf{\textcolor[rgb]{.933,0,0}{0.941}} &
    \textbf{\textcolor[rgb]{.933,0,0}{0.948}} &
    \textbf{\textcolor[rgb]{.933,0,0}{0.937}} &
    \textbf{\textcolor[rgb]{.933,0,0}{0.945}} &
    \textbf{\textcolor[rgb]{.933,0,0}{0.925}} &
    \textbf{\textcolor[rgb]{.933,0,0}{0.929}} \\
    \bottomrule
\end{tabular*}

\end{table*}

\begin{table*}[htbp]
\centering
\caption{Component-wise ablation of Life-IQA$^{+}$ on four IQA benchmark datasets.}
\label{table_7}
\footnotesize 
\setlength{\tabcolsep}{4.0pt}
\renewcommand{\arraystretch}{1.05}
\begin{tabular*}{0.78\textwidth}{@{\extracolsep{\fill}}l*{8}{c}@{}}
\toprule
    \multicolumn{1}{c}{Test on} &
    \multicolumn{2}{c}{LIVE} &
    \multicolumn{2}{c}{TID2013} &
    \multicolumn{2}{c}{KonIQ-10K} &
    \multicolumn{2}{c}{SPAQ} \\
    \cmidrule(lr){2-3}
    \cmidrule(lr){4-5}
    \cmidrule(lr){6-7}
    \cmidrule(lr){8-9}
    \multicolumn{1}{c}{Method} &
    \multicolumn{1}{c}{SRCC} & \multicolumn{1}{c}{PLCC} &
    \multicolumn{1}{c}{SRCC} & \multicolumn{1}{c}{PLCC} &
    \multicolumn{1}{c}{SRCC} & \multicolumn{1}{c}{PLCC} &
    \multicolumn{1}{c}{SRCC} & \multicolumn{1}{c}{PLCC} \\
    \midrule
    w/o GCN-enhanced interaction &
    0.977 & 0.979 &
    0.929 & 0.940 &
    0.928 & 0.938 &
    0.921 & 0.926 \\
    w/o stage3 &
    0.978 & 0.980 &
    0.922 & 0.934 &
    0.925 & 0.935 &
    0.923 & 0.926 \\
    w/o shared experts &
    0.965 & 0.972 &
    0.931 & 0.938 &
    0.934 & 0.943 &
    0.923 & 0.927 \\
    w/o routed experts &
    0.975 & 0.977 &
    0.937 & 0.946 &
    0.927 & 0.935 &
    0.922 & 0.927 \\
    w/o auxiliary task &
    0.978 & 0.979 &
    0.921 & 0.927 &
    0.934 & 0.943 &
    0.922 & 0.927 \\
    w/o rank loss &
    0.980 & 0.980 &
    0.930 & 0.937 &
    0.926 & 0.936 &
    0.921 & 0.926 \\
    \midrule
    Life-IQA$^{+}$ &
    \textbf{\textcolor[rgb]{.933,0,0}{0.983}} &
    \textbf{\textcolor[rgb]{.933,0,0}{0.984}} &
    \textbf{\textcolor[rgb]{.933,0,0}{0.941}} &
    \textbf{\textcolor[rgb]{.933,0,0}{0.948}} &
    \textbf{\textcolor[rgb]{.933,0,0}{0.937}} &
    \textbf{\textcolor[rgb]{.933,0,0}{0.945}} &
    \textbf{\textcolor[rgb]{.933,0,0}{0.925}} &
    \textbf{\textcolor[rgb]{.933,0,0}{0.929}} \\
    \bottomrule
\end{tabular*}

\end{table*}

\subsection{Effect of Auxiliary Supervision Placement}
To investigate how the placement of auxiliary classification affects quality prediction, we compare several variants with different auxiliary supervision configurations, as summarized in Table~\ref{table_9}. \textbf{\emph{No Auxiliary Supervision}} removes the auxiliary classification objective from Life-IQA$^{+}$, serving as the baseline without auxiliary supervision. \emph{Layer Interaction Supervision} uses the output of the GCN-enhanced layer interaction module, before hierarchical MoE processing, for auxiliary classification. \textbf{\emph{Hierarchical MoE Supervision}} uses the final output of the hierarchical MoE for auxiliary classification. \textbf{\emph{All Shared Experts Supervision}} uses the outputs of all shared experts for auxiliary classification. We also include \textbf{\emph{Transformer Decoder Supervision}} as an architectural baseline, which replaces all Life-IQA$^{+}$ modules following backbone feature extraction with a complete Transformer decoder and uses its output for auxiliary classification. In the full Life-IQA$^{+}$, only the outputs of the designated auxiliary shared experts are used for auxiliary classification, while the general shared experts do not receive direct auxiliary supervision. These comparisons assess whether selectively supervising designated shared experts is more effective for quality prediction than applying auxiliary classification at other locations or to all shared experts.

Compared with the no-auxiliary baseline, \emph{Layer Interaction Supervision} improves SRCC on TID2013 and KonIQ-10K but reduces it on LIVE. \emph{Hierarchical MoE Supervision} improves LIVE and KonIQ-10K while matching the baseline on TID2013. \emph{All Shared Experts Supervision} reduces SRCC on LIVE and TID2013 while leaving KonIQ-10K unchanged. \emph{Transformer Decoder Supervision} improves LIVE but substantially degrades TID2013 and KonIQ-10K. The full Life-IQA$^{+}$ achieves the highest SRCC on all three datasets, reaching 0.983 on LIVE, 0.941 on TID2013, and 0.937 on KonIQ-10K. These results support applying auxiliary classification only to designated shared experts.

\subsection{Ablation Study}

\noindent\textbf{Hyperparameter Analysis.}
Table~\ref{table_10} examines the effects of expert count and layer depth on CSIQ, LIVE, and LIVEC. Increasing the number of experts from four to eight improves SRCC across all three datasets, with gains from 0.959 to 0.970 on CSIQ and from 0.883 to 0.893 on LIVEC. Eight experts achieve the highest SRCC among the tested expert counts on all three datasets. Regarding layer depth, four layers achieve the highest SRCC on CSIQ and LIVE, reaching 0.970 and 0.983, respectively, while eight layers yield the highest SRCC of 0.896 on LIVEC. These results show distinct performance trends for expert count and layer depth: eight experts consistently provide the best results, whereas the benefit of additional layers depends on the dataset.

\noindent\textbf{Decoder Variant Comparison.}
Table~\ref{table_6} compares Life-IQA$^{+}$ with four alternative decoder configurations under the same training and evaluation protocol, all using four decoder layers. ``SwinT+Decoder'' uses a standard Transformer decoder, while ``SwinT+Parallel MoE'' organizes shared and routed experts in parallel. ``GCN$\rightarrow$MHA'' and ``MoE$\rightarrow$FFN'' replace the GCN-enhanced interaction with vanilla multi-head attention and the hierarchical MoE with a standard FFN, respectively, with all remaining components unchanged. Life-IQA$^{+}$ achieves the highest SRCC among the compared variants on all four datasets. Compared with ``GCN$\rightarrow$MHA'', it improves SRCC from 0.930 to 0.941 on TID2013. It also improves SPAQ SRCC from 0.920 to 0.925 compared with ``MoE$\rightarrow$FFN'', and KonIQ-10K SRCC from 0.933 to 0.937 compared with ``SwinT+Parallel MoE''. With decoder depth held constant, these gains support the contributions of GCN-enhanced feature interaction and hierarchical expert organization to quality prediction.

\noindent\textbf{Component-wise Ablation.}
Table~\ref{table_7} evaluates six variants, each removing one component while retaining the remaining architecture and training settings. The complete model achieves higher SRCC than every ablated variant on all four datasets. On TID2013, it achieves 0.941, compared with 0.929 without GCN-enhanced interaction and 0.922 without $stage3$ features. Shared and routed experts also contribute to prediction accuracy: the complete model improves LIVE SRCC from 0.965 to 0.983 compared with the variant without shared experts, and KonIQ-10K SRCC from 0.927 to 0.937 compared with the variant without routed experts. For the training objectives, TID2013 SRCC reaches 0.941 with both objectives enabled, compared with 0.921 without auxiliary supervision and 0.930 without the ranking loss. Together, these results support the contributions of cross-layer interaction, shared and routed expert processing, and both training objectives to the performance of Life-IQA$^{+}$.

\section{Conclusion}

In this paper, we propose Life-IQA$^{+}$, an effective quality representation decoding framework for blind image quality assessment. 
Instead of indiscriminately fusing multi-scale features, Life-IQA$^{+}$ focuses on selective interaction between deeper quality-aware representations. 
The GCN-enhanced layer interaction module uses the deepest-stage features to generate quality-oriented queries and aggregates complementary contextual information from the penultimate-stage features through cross-attention.

To further improve expert specialization for BIQA, we designed a hierarchical mixture-of-experts module. 
The shared expert layer learns degradation-aware quality priors with auxiliary IQA-related supervision, while the sparse routed expert layer dynamically selects a small subset of experts for adaptive quality refinement. 
Extensive experiments on multiple synthetic and authentic IQA benchmarks demonstrate that Life-IQA$^{+}$ achieves strong prediction accuracy, favorable model efficiency, and robust generalization. 
These findings suggest that task-oriented quality decoding and compact expert specialization are promising directions for future BIQA model design, especially under limited subjective annotations and diverse real-world degradation conditions.


\bibliographystyle{IEEEtran}
\bibliography{REFERENCES}

@INPROCEEDINGS{9156687,
  author={Su, Shaolin and Yan, Qingsen and Zhu, Yu and Zhang, Cheng and Ge, Xin and Sun, Jinqiu and Zhang, Yanning},
  booktitle = {Proceedings of the IEEE/CVF Conference on Computer Vision and Pattern Recognition},
  title={Blindly Assess Image Quality in the Wild Guided by a Self-Adaptive Hyper Network}, 
  year={2020},
  volume={},
  number={},
  pages={3664-3673},
  doi={10.1109/CVPR42600.2020.00372}}

@article{9817377,
author = {Pan, Zhaoqing and Zhang, Hao and Lei, Jianjun and Fang, Yuming and Shao, Xiao and Ling, Nam and Kwong, Sam},
title = {DACNN: Blind Image Quality Assessment via a Distortion-Aware Convolutional Neural Network},
journal = {IEEE Transactions on Circuits and Systems for Video Technology},
volume = 32,
number = 11,
pages = {7518-7531},
year = 2022
}

@article{9121773,
  author={Wu, Jinjian and Ma, Jupo and Liang, Fuhu and Dong, Weisheng and Shi, Guangming and Lin, Weisi},
  journal={IEEE Transactions on Image Processing}, 
  title={End-to-End Blind Image Quality Prediction With Cascaded Deep Neural Network}, 
  year={2020},
  volume={29},
  number={},
  pages={7414-7426},
}

@article{10109108,
  author={Sun, Wei and Min, Xiongkuo and Tu, Danyang and Ma, Siwei and Zhai, Guangtao},
  journal={IEEE Journal of Selected Topics in Signal Processing}, 
  title={Blind Quality Assessment for in-the-Wild Images via Hierarchical Feature Fusion and Iterative Mixed Database Training}, 
  year={2023},
  volume={17},
  number={6},
  pages={1178-1192},
}

@article{he2015deepresiduallearningimage,
  author    = {Kaiming He and Xiangyu Zhang and Shaoqing Ren and Jian Sun},
  title     = {Deep Residual Learning for Image Recognition},
  journal   = {arXiv preprint arXiv:1512.03385},
  year      = {2015}
}

@article{tan2020efficientnetrethinkingmodelscaling,
  author  = {Mingxing Tan and Quoc V. Le},
  title   = {{EfficientNet}: Rethinking Model Scaling for Convolutional Neural Networks},
  journal = {arXiv preprint arXiv:1905.11946},
  year    = {2020}
}

@article{howard2019searchingmobilenetv3,
  author    = {Andrew Howard and Mark Sandler and Grace Chu and Liang-Chieh Chen and Bo Chen and Mingxing Tan and Weijun Wang and Yukun Zhu and Ruoming Pang and Vijay Vasudevan and Quoc V. Le and Hartwig Adam},
  title     = {Searching for MobileNetV3},
  journal   = {arXiv preprint arXiv:1905.02244},
  year      = {2019}
}

@article{liu2021swintransformerhierarchicalvision,
  author    = {Ze Liu and Yutong Lin and Yue Cao and Han Hu and Yixuan Wei and Zheng Zhang and Stephen Lin and Baining Guo},
  title     = {Swin Transformer: Hierarchical Vision Transformer using Shifted Windows},
  journal   = {arXiv preprint arXiv:2103.14030},
  year      = {2021}
}

@article{tu2022maxvitmultiaxisvisiontransformer,
  author    = {Zhengzhong Tu and Hossein Talebi and Han Zhang and Feng Yang and Peyman Milanfar and Alan Bovik and Yinxiao Li},
  title     = {MaxViT: Multi-Axis Vision Transformer},
  journal   = {arXiv preprint arXiv:2204.01697},
  year      = {2022}
}

@article{wang2021pyramidvisiontransformerversatile,
  author    = {Wenhai Wang and Enze Xie and Xiang Li and Deng-Ping Fan and Kaitao Song and Ding Liang and Tong Lu and Ping Luo and Ling Shao},
  title     = {Pyramid Vision Transformer: A Versatile Backbone for Dense Prediction without Convolutions},
  journal   = {arXiv preprint arXiv:2102.12122},
  year      = {2021}
}

@ARTICLE{7327186,
  author={Ghadiyaram, Deepti and Bovik, Alan C.},
  journal={IEEE Transactions on Image Processing}, 
  title={Massive Online Crowdsourced Study of Subjective and Objective Picture Quality}, 
  year={2016},
  volume={25},
  number={1},
  pages={372-387},
  doi={10.1109/TIP.2015.2500021}}

@ARTICLE{9694502,
  author={Pan, Zhaoqing and Yuan, Feng and Lei, Jianjun and Fang, Yuming and Shao, Xiao and Kwong, Sam},
  journal={IEEE Transactions on Image Processing}, 
  title={VCRNet: Visual Compensation Restoration Network for No-Reference Image Quality Assessment}, 
  year={2022},
  volume={31},
  number={},
  pages={1613-1627},
}

@ARTICLE{10092271,
  author={Lan, Xuting and Zhou, Mingliang and Xu, Xueyong and Wei, Xuekai and Liao, Xingran and Pu, Huayan and Luo, Jun and Xiang, Tao and Fang, Bin and Shang, Zhaowei},
  journal={IEEE Transactions on Broadcasting}, 
  title={Multilevel Feature Fusion for End-to-End Blind Image Quality Assessment}, 
  year={2023},
  volume={69},
  number={3},
  pages={801-811},
  }

@INPROCEEDINGS{9706735,
  author={Golestaneh, S. Alireza and Dadsetan, Saba and Kitani, Kris M.},
  booktitle={Proceedings of the IEEE Winter Conference on Applications of Computer Vision}, 
  title={No-Reference Image Quality Assessment via Transformers, Relative Ranking, and Self-Consistency}, 
  year={2022},
  volume={},
  number={},
  pages={3989-3999},
}

@ARTICLE{10706861,
  author={Tang, Long and Han, Yongming and Yuan, Liang and Zhai, Guangtao},
  journal={IEEE Signal Processing Letters}, 
  title={FsPN: Blind Image Quality Assessment Based on Feature-Selected Pyramid Network}, 
  year={2025},
  volume={32},
  number={},
  pages={1-5},
}

@INPROCEEDINGS{8743252,
  author={Lin, Hanhe and Hosu, Vlad and Saupe, Dietmar},
  booktitle={Proceedings of the IEEE International Conference on Quality of Multimedia Experience}, 
  title={KADID-10k: A Large-scale Artificially Distorted IQA Database}, 
  year={2019},
  volume={},
  number={},
  pages={1-3},
  doi={10.1109/QoMEX.2019.8743252}}

@INPROCEEDINGS{9506075,
  author={You, Junyong and Korhonen, Jari},
  booktitle={Proceedings of the IEEE International Conference on Image Processing}, 
  title={Transformer For Image Quality Assessment}, 
  year={2021},
  volume={},
  number={},
  pages={1389-1393},
}

@INPROCEEDINGS{9710973,
  author={Ke, Junjie and Wang, Qifei and Wang, Yilin and Milanfar, Peyman and Yang, Feng},
  booktitle={Proceedings of the IEEE/CVF International Conference on Computer Vision}, 
  title={MUSIQ: Multi-scale Image Quality Transformer}, 
  year={2021},
  volume={},
  number={},
  pages={5128-5137},
}

@ARTICLE{6797059,
  author={Jacobs, Robert A. and Jordan, Michael I. and Nowlan, Steven J. and Hinton, Geoffrey E.},
  journal={Neural Computation}, 
  title={Adaptive Mixtures of Local Experts}, 
  year={1991},
  volume={3},
  number={1},
  pages={79-87},
 }

@article{lepikhin2020gshardscalinggiantmodels,
  author    = {Dmitry Lepikhin and HyoukJoong Lee and Yuanzhong Xu and Dehao Chen and Orhan Firat and Yanping Huang and Maxim Krikun and Noam Shazeer and Zhifeng Chen},
  title     = {GShard: Scaling Giant Models with Conditional Computation and Automatic Sharding},
  journal   = {arXiv preprint arXiv:2006.16668},
  year      = {2020}
}

@article{jiang2024mixtralexperts,
  author    = {Albert Q. Jiang and Alexandre Sablayrolles and Antoine Roux and Arthur Mensch and Blanche Savary and Chris Bamford and Devendra Singh Chaplot and Diego de las Casas and Emma Bou Hanna and Florian Bressand and Gianna Lengyel and Guillaume Bour and Guillaume Lample and Lélio Renard Lavaud and Lucile Saulnier and Marie-Anne Lachaux and Pierre Stock and Sandeep Subramanian and Sophia Yang and Szymon Antoniak and Teven Le Scao and Théophile Gervet and Thibaut Lavril and Thomas Wang and Timothée Lacroix and William El Sayed},
  title     = {Mixtral of Experts},
  journal   = {arXiv preprint arXiv:2401.04088},
  year      = {2024}
}

@article{dai2024deepseekmoeultimateexpertspecialization,
  author  = {Damai Dai and Chengqi Deng and Chenggang Zhao and
             R. X. Xu and Huazuo Gao and Deli Chen and Jiashi Li and
             Wangding Zeng and Xingkai Yu and Y. Wu and Zhenda Xie and
             Y. K. Li and Panpan Huang and Fuli Luo and Chong Ruan and
             Zhifang Sui and Wenfeng Liang},
  title   = {{DeepSeekMoE}: Towards Ultimate Expert Specialization in
             Mixture-of-Experts Language Models},
  journal = {arXiv preprint arXiv:2401.06066},
  year    = {2024}
}

@article{qin2023dataefficientimagequalityassessment,
  author    = {Guanyi Qin and Runze Hu and Yutao Liu and Xiawu Zheng and Haotian Liu and Xiu Li and Yan Zhang},
  title     = {Data-Efficient Image Quality Assessment with Attention-Panel Decoder},
  journal   = {arXiv preprint arXiv:2304.04952},
  year      = {2023}
}

@ARTICLE{1709988,
  author={Sheikh, H.R. and Sabir, M.F. and Bovik, A.C.},
  journal={IEEE Transactions on Image Processing}, 
  title={A Statistical Evaluation of Recent Full Reference Image Quality Assessment Algorithms}, 
  year={2006},
  volume={15},
  number={11},
  pages={3440-3451},
  doi={10.1109/TIP.2006.881959}}

@article{larson2010most,
  title={Most apparent distortion: full-reference image quality assessment and the role of strategy},
  author={Larson, Eric C and Chandler, Damon M},
  journal={Journal of electronic imaging},
  volume={19},
  number={1},
  pages={011006--011006},
  year={2010},
  publisher={Society of Photo-Optical Instrumentation Engineers}
}

@article{PONOMARENKO201557,
title = {Image database TID2013: Peculiarities, results and perspectives},
journal = {Signal Processing: Image Communication},
volume = {30},
pages = {57-77},
year = {2015},
issn = {0923-5965},
doi = {https://doi.org/10.1016/j.image.2014.10.009},
author = {Nikolay Ponomarenko and Lina Jin and Oleg Ieremeiev and Vladimir Lukin and Karen Egiazarian and Jaakko Astola and Benoit Vozel and Kacem Chehdi and Marco Carli and Federica Battisti and C.-C. {Jay Kuo}},
}

@INPROCEEDINGS{9156490,
  author={Fang, Yuming and Zhu, Hanwei and Zeng, Yan and Ma, Kede and Wang, Zhou},
  booktitle={Proceedings of the IEEE/CVF Conference on Computer Vision and Pattern Recognition}, 
  title={Perceptual Quality Assessment of Smartphone Photography}, 
  year={2020},
  volume={},
  number={},
  pages={3674-3683},
}

@ARTICLE{8968750,
  author={Hosu, Vlad and Lin, Hanhe and Sziranyi, Tamas and Saupe, Dietmar},
  journal={IEEE Transactions on Image Processing}, 
  title={KonIQ-10k: An Ecologically Valid Database for Deep Learning of Blind Image Quality Assessment}, 
  year={2020},
  volume={29},
  number={},
  pages={4041-4056},
  doi={10.1109/TIP.2020.2967829}}

@ARTICLE{10213444,
  author={Hu, Bo and Zhu, Guang and Li, Leida and Gan, Ji and Li, Weisheng and Gao, Xinbo},
  journal={IEEE Transactions on Multimedia}, 
  title={Blind Image Quality Index With Cross-Domain Interaction and Cross-Scale Integration}, 
  year={2024},
  volume={26},
  number={},
  pages={2729-2739},
  doi={10.1109/TMM.2023.3303725}}

@ARTICLE{10520932,
  author={Li, Aobo and Wu, Jinjian and Liu, Yongxu and Li, Leida and Dong, Weisheng and Shi, Guangming},
  journal={IEEE Transactions on Multimedia}, 
  title={Blind Image Quality Assessment Based on Perceptual Comparison}, 
  year={2024},
  volume={},
  number={},
  pages={1-12},
  doi={10.1109/TMM.2024.3397051}}

@ARTICLE{10287911,
  author={Shi, Jinsong and Gao, Pan and Smolic, Aljosa},
  journal={IEEE Transactions on Multimedia}, 
  title={Blind Image Quality Assessment via Transformer Predicted Error Map and Perceptual Quality Token}, 
  year={2024},
  volume={26},
  number={},
  pages={4641-4651},
}

@INPROCEEDINGS{10204315,
  author={Zhang, Weixia and Zhai, Guangtao and Wei, Ying and Yang, Xiaokang and Ma, Kede},
  booktitle={Proceedings of the IEEE/CVF Conference on Computer Vision and Pattern Recognition}, 
  title={Blind Image Quality Assessment via Vision-Language Correspondence: A Multitask Learning Perspective}, 
  year={2023},
  volume={},
  number={},
  pages={14071-14081},
  doi={10.1109/CVPR52729.2023.01352}}

@INPROCEEDINGS{10204004,
  author={Saha, Avinab and Mishra, Sandeep and Bovik, Alan C.},
  booktitle={Proceedings of the IEEE/CVF Conference on Computer Vision and Pattern Recognition}, 
  title={Re-IQA: Unsupervised Learning for Image Quality Assessment in the Wild}, 
  year={2023},
  volume={},
  number={},
  pages={5846-5855},
  doi={10.1109/CVPR52729.2023.00566}}

@INPROCEEDINGS{10656546,
  author={Shin, Nyeong-Ho and Lee, Seon-Ho and Kim, Chang-Su},
  booktitle={Proceedings of the IEEE/CVF Conference on Computer Vision and Pattern Recognition}, 
  title={Blind Image Quality Assessment Based on Geometric Order Learning}, 
  year={2024},
  volume={},
  number={},
  pages={12799-12808},
  doi={10.1109/CVPR52733.2024.01216}}

@INPROCEEDINGS{10655501,
  author={Xu, Kangmin and Liao, Liang and Xiao, Jing and Chen, Chaofeng and Wu, Haoning and Yan, Qiong and Lin, Weisi},
  booktitle={Proceedings of the IEEE/CVF Conference on Computer Vision and Pattern Recognition}, 
  title={Boosting Image Quality Assessment Through Efficient Transformer Adaptation with Local Feature Enhancement}, 
  year={2024},
  volume={},
  number={},
  pages={2662-2672},
  doi={10.1109/CVPR52733.2024.00257}}

@inproceedings{10222634,
  author    = {Zesheng Wang and Wei Wu and Liang Yuan and Wei Sun and Ying Chen and Kai Li and Guangtao Zhai},
  title     = {Hierarchical Feature Fusion Transformer for No-Reference Image Quality Assessment},
  booktitle = {Proceedings of the IEEE International Conference on Image Processing},
  year      = {2023},
  pages     = {2205--2209},
  doi       = {10.1109/ICIP49359.2023.10222634}
}

@ARTICLE{9337868,
  author={Li, Fan and Zhang, Yangfan and Cosman, Pamela C.},
  journal={IEEE Transactions on Circuits and Systems for Video Technology}, 
  title={MMMNet: An End-to-End Multi-Task Deep Convolution Neural Network With Multi-Scale and Multi-Hierarchy Fusion for Blind Image Quality Assessment}, 
  year={2021},
  volume={31},
  number={12},
  pages={4798-4811},
  doi={10.1109/TCSVT.2021.3055197}}

@ARTICLE{10845143,
  author={Shi, Haozhi and Xie, Weiying and Qin, Haonan and Li, Yunsong and Fang, Leyuan},
  journal={IEEE Transactions on Circuits and Systems for Video Technology}, 
  title={Visual State Space Model With Graph-Based Feature Aggregation for No-Reference Image Quality Assessment}, 
  year={2025},
  volume={35},
  number={6},
  pages={5589-5601},
  doi={10.1109/TCSVT.2025.3527930}}

@INPROCEEDINGS{11092710,
  author={Li, Xudong and Nie, Wenjie and Zhang, Yan and Hu, Runze and Li, Ke and Zheng, Xiawu and Cao, Liujuan},
  booktitle={Proceedings of the IEEE/CVF Conference on Computer Vision and Pattern Recognition}, 
  title={Distilling Spatially-Heterogeneous Distortion Perception for Blind Image Quality Assessment}, 
  year={2025},
  volume={},
  number={},
  pages={2344-2354},

  doi={10.1109/CVPR52734.2025.00224}}

@INPROCEEDINGS{11093358,
  author={Shen, Wenhao and Zhou, Mingliang and Chen, Yu and Wei, Xuekai and Feng, Yong and Pu, Huayan and Jia, Weijia},
  booktitle={Proceedings of the IEEE/CVF Conference on Computer Vision and Pattern Recognition}, 
  title={Image Quality Assessment: Investigating Causal Perceptual Effects with Abductive Counterfactual Inference}, 
  year={2025},
  volume={},
  number={},
  pages={17990-17999},

  doi={10.1109/CVPR52734.2025.01676}}

@article{dosovitskiy2021imageworth16x16words,
  author    = {Alexey Dosovitskiy and Lucas Beyer and Alexander Kolesnikov and Dirk Weissenborn and Xiaohua Zhai and Thomas Unterthiner and Mostafa Dehghani and Matthias Minderer and Georg Heigold and Sylvain Gelly and Jakob Uszkoreit and Neil Houlsby},
  title     = {An Image is Worth 16x16 Words: Transformers for Image Recognition at Scale},
  journal   = {arXiv preprint arXiv:2010.11929},
  year      = {2021}
}

@ARTICLE{10960664,
  author={Feng, Chenxi and Min, Xiongkuo and Ye, Long and Yang, Yinghao},
  journal={IEEE Transactions on Circuits and Systems for Video Technology}, 
  title={LCIQA: A Lightweight Contrastive-Learning Framework for Image Quality Assessment via Cross-Scale Consistency Minimization}, 
  year={2025},
  volume={35},
  number={10},
  pages={9986-9999},
  doi={10.1109/TCSVT.2025.3558797}}

@inproceedings{10647859,
  author    = {Long Tang and Liang Yuan and Guoquan Zheng and Zesheng Wang and Guangtao Zhai},
  title     = {DTSN: No-Reference Image Quality Assessment via Deformable Transformer and Semantic Network},
  booktitle = {Proceedings of the IEEE International Conference on Image Processing},
  year      = {2024},
  pages     = {1207--1211},
  doi       = {10.1109/ICIP51287.2024.10647859}
}

@ARTICLE{10833846,
  author={Liao, Xingran and Wei, Xuekai and Zhou, Mingliang and Wong, Hau-San and Kwong, Sam},
  journal={IEEE Transactions on Pattern Analysis and Machine Intelligence}, 
  title={Image Quality Assessment: Exploring Joint Degradation Effect of Deep Network Features via Kernel Representation Similarity Analysis}, 
  year={2025},
  volume={47},
  number={4},
  pages={2799-2815},
  doi={10.1109/TPAMI.2025.3527004}}

@ARTICLE{10478301,
  author={Chen, Chaofeng and Mo, Jiadi and Hou, Jingwen and Wu, Haoning and Liao, Liang and Sun, Wenxiu and Yan, Qiong and Lin, Weisi},
  journal={IEEE Transactions on Image Processing}, 
  title={TOPIQ: A Top-Down Approach From Semantics to Distortions for Image Quality Assessment}, 
  year={2024},
  volume={33},
  number={},
  pages={2404-2418},
  doi={10.1109/TIP.2024.3378466}}

@ARTICLE{10795128,
  author={Zhou, Fei and Gu, Tianhao and Huang, Zhicong and Qiu, Guoping},
  journal={IEEE Journal of Selected Topics in Signal Processing}, 
  title={Vision Language Modeling of Content, Distortion and Appearance for Image Quality Assessment}, 
  year={2025},
  volume={19},
  number={1},
  pages={234-247},
  doi={10.1109/JSTSP.2024.3516392}}

@ARTICLE{11175315,
  author={Sang, Qingbing and Li, Qian and Liu, Lixiong and Deng, Zhaohong and Wu, Xiaojun and Bovik, Alan C.},
  journal={IEEE Transactions on Image Processing}, 
  title={No-Reference Image Quality Assessment Leveraging GenAI Images}, 
  year={2025},
  volume={34},
  number={},
  pages={6204-6214},
  doi={10.1109/TIP.2025.3610238}}

@ARTICLE{10463077,
  author={Tian, Yu and Wang, Shiqi and Chen, Baoliang and Kwong, Sam},
  journal={IEEE Transactions on Circuits and Systems for Video Technology}, 
  title={Causal Representation Learning for GAN-Generated Face Image Quality Assessment}, 
  year={2024},
  volume={34},
  number={8},
  pages={7589-7600},
  doi={10.1109/TCSVT.2024.3374887}}

@ARTICLE{10462930,
  author={Zhang, Weixia and Ma, Kede and Zhai, Guangtao and Yang, Xiaokang},
  journal={IEEE Transactions on Image Processing}, 
  title={Task-Specific Normalization for Continual Learning of Blind Image Quality Models}, 
  year={2024},
  volume={33},
  number={},
  pages={1898-1910},
  doi={10.1109/TIP.2024.3371349}}

@ARTICLE{10323206,
  author={Ding, Keyan and Zhong, Rijin and Wang, Zhihua and Yu, Yang and Fang, Yuming},
  journal={IEEE Transactions on Multimedia}, 
  title={Adaptive Structure and Texture Similarity Metric for Image Quality Assessment and Optimization}, 
  year={2024},
  volume={26},
  number={},
  pages={5398-5409},
  doi={10.1109/TMM.2023.3333208}}

@ARTICLE{11226953,
  author={Han, Xumeng and Wei, Longhui and Dou, Zhiyang and Sun, Yingfei and Han, Zhenjun and Tian, Qi},
  journal={IEEE Transactions on Image Processing}, 
  title={ViMoE: An Empirical Study of Designing Vision Mixture-of-Experts}, 
  year={2025},
  volume={},
  number={},
  pages={1-1},
  doi={10.1109/TIP.2025.3626887}}

@article{zhou2023brainformers,
  author    = {Yanqi Zhou and Nan Du and Yanping Huang and Daiyi Peng and Chang Lan and Da Huang and Siamak Shakeri and David So and Andrew Dai and Yifeng Lu and Zhifeng Chen and Quoc Le and Claire Cui and James Laudon and Jeff Dean},
  title     = {Brainformers: Trading Simplicity for Efficiency},
  journal   = {arXiv preprint arXiv:2306.00008},
  year      = {2024}
}

@inproceedings{10.5555/3618408.3618649,
  author    = {Mohammed Nowaz Rabbani Chowdhury and Shuai Zhang and Meng Wang and Sijia Liu and Pin-Yu Chen},
  title     = {Patch-level routing in mixture-of-experts is provably sample-efficient for convolutional neural networks},
  booktitle = {Proceedings of the International Conference on Machine Learning},
  year      = {2023}
}

@ARTICLE{10056825,
  author={Liu, Manni and Huang, Jiabin and Zeng, Delu and Ding, Xinghao and Paisley, John},
  journal={IEEE Transactions on Image Processing}, 
  title={A Multiscale Approach to Deep Blind Image Quality Assessment}, 
  year={2023},
  volume={32},
  number={},
  pages={1656-1667},
  doi={10.1109/TIP.2023.3245991}}

@ARTICLE{10036133,
  author={Ma, Rui and Wu, Qingbo and Ngan, King Ngi and Li, Hongliang and Meng, Fanman and Xu, Linfeng},
  journal={IEEE Transactions on Multimedia}, 
  title={Forgetting to Remember: A Scalable Incremental Learning Framework for Cross-Task Blind Image Quality Assessment}, 
  year={2023},
  volume={25},
  number={},
  pages={8817-8827},
  doi={10.1109/TMM.2023.3242143}}

@ARTICLE{9961939,
  author={Zhu, Yucheng and Li, Yunhao and Sun, Wei and Min, Xiongkuo and Zhai, Guangtao and Yang, Xiaokang},
  journal={IEEE Transactions on Multimedia}, 
  title={Blind Image Quality Assessment via Cross-View Consistency}, 
  year={2023},
  volume={25},
  number={},
  pages={7607-7620},
  doi={10.1109/TMM.2022.3224319}}

@ARTICLE{9720138,
  author={Sun, Simeng and Yu, Tao and Xu, Jiahua and Zhou, Wei and Chen, Zhibo},
  journal={IEEE Transactions on Multimedia}, 
  title={GraphIQA: Learning Distortion Graph Representations for Blind Image Quality Assessment}, 
  year={2023},
  volume={25},
  number={},
  pages={2912-2925},
  doi={10.1109/TMM.2022.3152942}}

@inproceedings{7351311,
  author={Kang, Le and Ye, Peng and Li, Yi and Doermann, David},
  booktitle={Proceedings of the IEEE International Conference on Image Processing }, 
  title={Simultaneous estimation of image quality and distortion via multi-task convolutional neural networks}, 
  year={2015},
  volume={},
  number={},
  pages={2791-2795},
  doi={10.1109/ICIP.2015.7351311}}

@ARTICLE{8576582,
  author={Zhang, Weixia and Ma, Kede and Yan, Jia and Deng, Dexiang and Wang, Zhou},
  journal={IEEE Transactions on Circuits and Systems for Video Technology}, 
  title={Blind Image Quality Assessment Using a Deep Bilinear Convolutional Neural Network}, 
  year={2020},
  volume={30},
  number={1},
  pages={36-47},
  doi={10.1109/TCSVT.2018.2886771}}

@inproceedings{7780802,
  author={Misra, Ishan and Shrivastava, Abhinav and Gupta, Abhinav and Hebert, Martial},
  booktitle={Proceedings of the IEEE Conference on Computer Vision and Pattern Recognition}, 
  title={Cross-Stitch Networks for Multi-task Learning}, 
  year={2016},
  volume={},
  number={},
  pages={3994-4003},
  doi={10.1109/CVPR.2016.433}}

@inproceedings{Tang_2026_CVPR,
    author    = {Tang, Long and Duan, Huiyu and Zheng, Guoquan and Zhang, Jianbo and Hao, Jie and Yuan, Liang},
    title     = {Life-IQA: Boosting Blind Image Quality Assessment through GCN-enhanced Layer Interaction and MoE-based Feature Decoupling},
    booktitle = {Proceedings of the IEEE/CVF Conference on Computer Vision and Pattern Recognition },
    month     = {June},
    year      = {2026},
    pages     = {29866-29876}
}

@article{wu2023qalign,
  author  = {Haoning Wu and Zicheng Zhang and Weixia Zhang and
             Chaofeng Chen and Chunyi Li and Liang Liao and Annan Wang and
             Erli Zhang and Wenxiu Sun and Qiong Yan and Xiongkuo Min and
             Guangtao Zhai and Weisi Lin},
  title   = {{Q-Align}: Teaching {LMMs} for Visual Scoring via
             Discrete Text-Defined Levels},
  journal = {arXiv preprint arXiv:2312.17090},
  year    = {2023}
}

@InProceedings{Duan_2025_CVPR,
    author    = {Duan, Huiyu and Hu, Qiang and Wang, Jiarui and Yang, Liu and Xu, Zitong and Liu, Lu and Min, Xiongkuo and Cai, Chunlei and Ye, Tianxiao and Zhang, Xiaoyun and Zhai, Guangtao},
    title     = {FineVQ: Fine-Grained User Generated Content Video Quality Assessment},
    booktitle = {Proceedings of the Computer Vision and Pattern Recognition Conference},
    month     = {June},
    year      = {2025},
    pages     = {3206-3217}
}

@article{eigen2014learningfactoredrepresentationsdeep,
  author  = {David Eigen and Marc'Aurelio Ranzato and Ilya Sutskever},
  title   = {Learning Factored Representations in a Deep Mixture of Experts},
  journal = {arXiv preprint arXiv:1312.4314},
  year    = {2014}
}

\newpage
\begin{IEEEbiography}[{\includegraphics[width=1in,height=1.25in,clip,keepaspectratio]{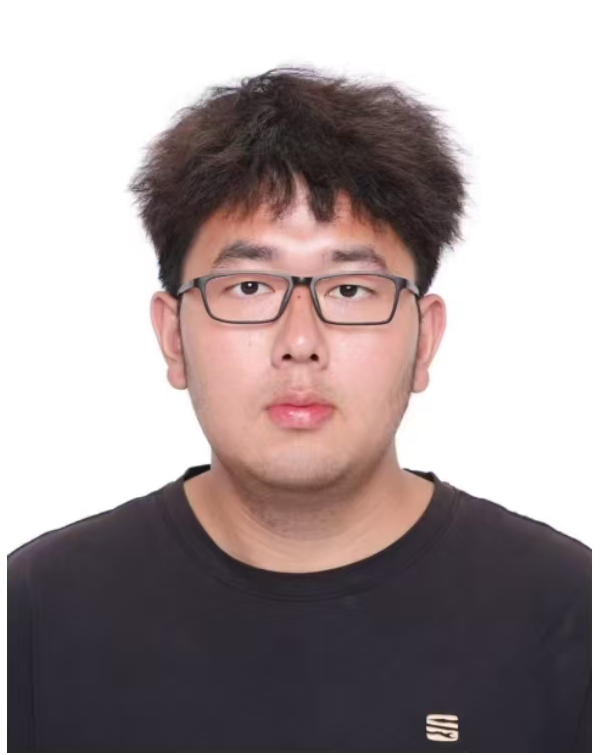}}]{Long Tang}
received the bachelor’s degree in information science and technology and the M.S. degree in control science and engineering from Beijing University of Chemical Technology. He is currently pursuing the Ph.D. degree at the School of Integrated Circuits, Shanghai Jiao Tong University. His research interests include perceptual image quality assessment and image processing.
\end{IEEEbiography}

\begin{IEEEbiography}[{\includegraphics[width=1in,height=1.25in,clip,keepaspectratio]{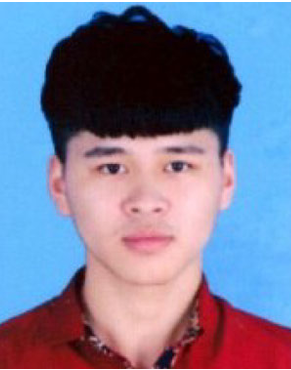}}]{Guoquan Zheng}
received the master's degree in mechanical engineering from Xinjiang University, Urumqi, China. He is currently pursuing the Ph.D. degree with the School of Information Science and Technology, Beijing University of Chemical Technology, Beijing, China. His current research interests include 3D human body reconstruction and quality assessment.
\end{IEEEbiography}

\begin{IEEEbiography}[{\includegraphics[width=1in,height=1.25in,clip,keepaspectratio]{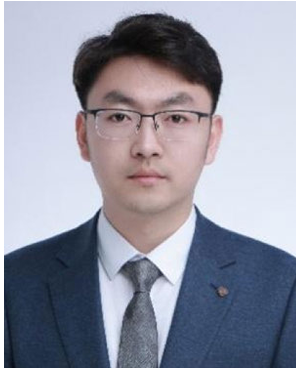}}]{Huiyu Duan}
received the B.E. degree from the University of Electronic Science and Technology of China, Chengdu, China, in 2017, and the Ph.D. degree from Shanghai Jiao Tong University, Shanghai, China, in 2024. From September 2019 to 2020, he was a Visiting Ph.D. Student with the Schepens Eye Research Institute, Harvard Medical School, Boston, USA. He is currently a Post-Doctoral Fellow with Shanghai Jiao Tong University. His research interests include perceptual quality assessment, quality of experience, visual attention modeling, extended reality (XR), and multimedia signal processing. He received the Best Paper Award of IEEE International Symposium on Broadband Multimedia Systems and Broadcasting (BMSB) in 2022.
\end{IEEEbiography}

\begin{IEEEbiography}[{\includegraphics[width=1in,height=1.25in,clip,keepaspectratio]{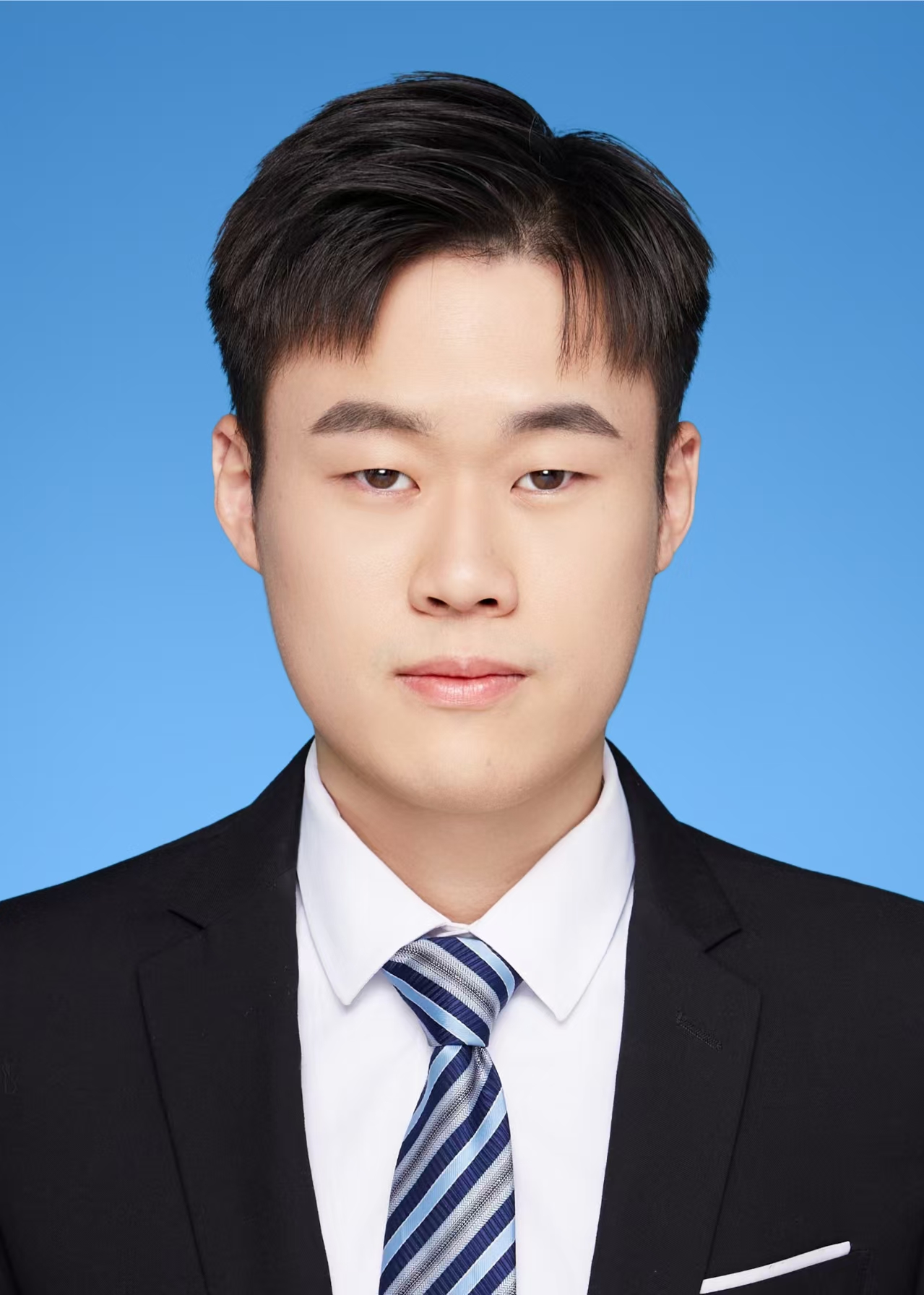}}]{Jie Hao}
received Master's degree from Beijing University of Chemical Technology in 2026 and is currently pursuing a Ph,D. degree at Shanghai Jiao Tong University. His current research interests include image quality assessment, quality assessment for novel view synthesis (NVS) scenes, and mesh quality assessment.
\end{IEEEbiography}

\begin{IEEEbiography}[{\includegraphics[width=1in,height=1.25in,clip,keepaspectratio]{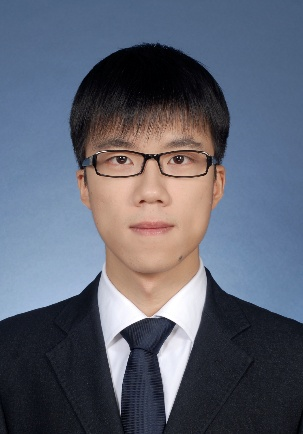}}]{Yucheng Zhu}
received the B.E. degree from the Shanghai Jiao Tong University, Shanghai, China, in 2015, and the Ph.D. degree from Shanghai Jiao Tong University, Shanghai, China, in 2021, where he is currently an associate professor with the SJTU Institute of Cultural and Creative Industry. His research interests include visual quality assessment, visual attention modeling and perceptual signal processing.
\end{IEEEbiography}

\begin{IEEEbiography}[{\includegraphics[width=1in,height=1.25in,clip,keepaspectratio]{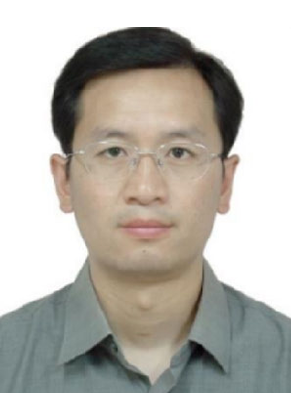}}]{Liang Yuan}
received the Ph.D. degree in electrical and computer engineering from The Ohio State University, Columbus, in 2011. He is currently a Professor of the USC-SJTU Institute of Cultural and Creative Industry, Shanghai Jiao Tong University, Shanghai. His research interests include computer vision and imaging processes, robotics, and automation in life science. He was honored with the National Recruitment Program of Global Experts in 2014.
\end{IEEEbiography}

\begin{IEEEbiography}[{\includegraphics[width=1in,height=1.25in,clip,keepaspectratio]{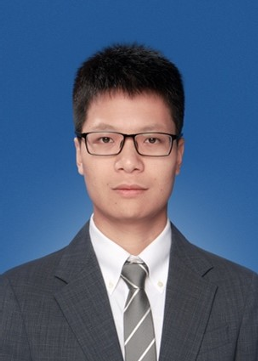}}]{Xiongkuo Min}
(Member, IEEE) received the B.E. degree from Wuhan University, Wuhan, China, in 2013, and the Ph.D. degree from Shanghai Jiao Tong University, Shanghai, China, in 2018, where he is currently a tenure-track Associate Professor with the Institute of Image Communication and Network Engineering. From Jun. 2018 to Sept. 2021, he was a Postdoc at Shanghai Jiao Tong University. From Jan. 2016 to Jan. 2017, he was a visiting student at University of Waterloo. From Jan. 2019 to Jan. 2021, he was a visiting scholar at The University of Texas at Austin and the University of Macau. He received the Best Paper Runner-up Award of IEEE Transactions on Multimedia in 2021, the Best Student Paper Award of IEEE International Conference on Multimedia and Expo (ICME) in 2016, the Best Paper Award of IEEE International Symposium on Broadband Multimedia Systems and Broadcasting (BMSB) in 2022, and several first place awards of grand challenges held at IEEE ICME and ICIP. His research interests include image/video/audio quality, quality of experience, multimedia, image/video processing, computer vision.
\end{IEEEbiography}

\begin{IEEEbiography}[{\includegraphics[width=1in,height=1.25in,clip,keepaspectratio]{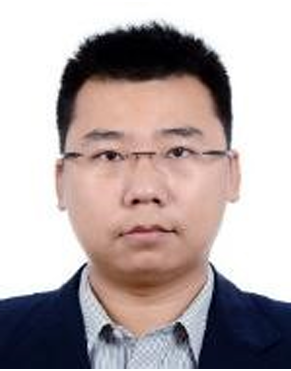}}]{Guangtao Zhai}
(Fellow, IEEE) received the B.E. and M.E. degrees from Shandong University, Shandong, China, in 2001 and 2004, respectively, and the Ph.D. degree from Shanghai Jiao Tong University, Shanghai, China, in 2009. From 2008 to 2009, he was a Visiting Student with the Department of Electrical and Computer Engineering, McMaster University, Hamilton, ON, Canada, where he was a Post-Doctoral Fellow from 2010 to 2012. From 2012 to 2013, he was a Humboldt Research Fellow with the Institute of Multimedia Communication and Signal Processing, Friedrich-Alexander University of Erlangen-Nuremberg, Germany. He is currently a Professor with the Department of Electronics Engineering, Shanghai Jiao Tong University.
\end{IEEEbiography}

\vfill

\end{document}